\documentclass[format=acmsmall, review=false, screen=true]{acmart}

\usepackage{booktabs}

\usepackage{etex}

\usepackage{subcaption}
\usepackage[font={normalsize}]{caption}
\usepackage{colortbl}

\usepackage{graphicx}
\usepackage{booktabs}
\usepackage{epstopdf}
\usepackage{url}
\usepackage{placeins}
\usepackage{amsmath,amsfonts}
\usepackage{mathtools,mathrsfs}
\usepackage{multirow}
\usepackage{rotating}
\usepackage{pbox}
\usepackage[normalem]{ulem}
\usepackage{listings}

\usepackage{cleveref}
\usepackage[utf8]{inputenc}
\crefname{section}{§}{§§}
\Crefname{section}{§}{§§}

\usepackage[10pt]{moresize}

\usepackage{xcolor}
\definecolor{darkgrey}{RGB}{70,70,70}
\definecolor{lightgrey}{RGB}{200,200,200}

\usepackage{soul}

\usepackage{enumitem}

\makeatletter
\newenvironment{btHighlight}[1][]
{\begingroup\tikzset{bt@Highlight@par/.style={#1}}\begin{lrbox}{\@tempboxa}}
{\end{lrbox}\bt@HL@box[bt@Highlight@par]{\@tempboxa}\endgroup}

\newcommand\btHL[1][]{%
  \begin{btHighlight}[#1]\bgroup\aftergroup\bt@HL@endenv%
}
\def\bt@HL@endenv{%
  \end{btHighlight}%
  \egroup
}
\newcommand{\bt@HL@box}[2][]{%
  \tikz[#1]{%
    \pgfpathrectangle{\pgfpoint{1pt}{0pt}}{\pgfpoint{\wd #2}{\ht #2}}%
    \pgfusepath{use as bounding box}%
    \node[anchor=base west, fill=orange!30,outer sep=0pt,inner xsep=1pt, inner ysep=0pt, rounded corners=3pt, minimum height=\ht\strutbox+1pt,#1]{\raisebox{1pt}{\strut}\strut\usebox{#2}};
  }%
}
\makeatother

\usepackage[linesnumbered,ruled,vlined]{algorithm2e}
\SetAlFnt{\scriptsize\tt}
\SetAlCapFnt{\scriptsize\sf}
\SetAlCapNameFnt{\scriptsize\sf}

\newcommand{\macb}[1]{\textbf{\textsf{#1}}}

\usepackage[hang,flushmargin]{footmisc}

\usepackage{bm}

\usepackage{scalerel,stackengine}
\stackMath
\newcommand\rwh[1]{%
\savestack{\tmpbox}{\stretchto{%
  \scaleto{%
      \scalerel*[\widthof{\ensuremath{#1}}]{\kern-.6pt\bigwedge\kern-.6pt}%
          {\rule[-\textheight/2]{1ex}{\textheight}}
            }{\textheight}%
}{0.5ex}}%
\stackon[1pt]{#1}{\tmpbox}%
}

\def\HiLiGA{\leavevmode\rlap{\hbox to \hsize{\color{black!10}\leaders\hrule height 1\baselineskip depth 1ex\hfill}}}
\def\HiLiGB{\leavevmode\rlap{\hbox to \hsize{\color{black!25}\leaders\hrule height 1\baselineskip depth 1ex\hfill}}}
\def\HiLiGC{\leavevmode\rlap{\hbox to \hsize{\color{black!40}\leaders\hrule height 1\baselineskip depth 1ex\hfill}}}
\def\HiLiGD{\leavevmode\rlap{\hbox to \hsize{\color{black!55}\leaders\hrule height 1\baselineskip depth 1ex\hfill}}}
\def\HiLiGE{\leavevmode\rlap{\hbox to \hsize{\color{black!70}\leaders\hrule height 1\baselineskip depth 1ex\hfill}}}
\def\HiLiGF{\leavevmode\rlap{\hbox to \hsize{\color{black!85}\leaders\hrule height 1\baselineskip depth 1ex\hfill}}}

\usepackage{algpseudocode}
\usepackage{tikz}
\usetikzlibrary{calc}
\usepackage{xcolor}
\makeatletter

\makeatletter 
\renewcommand{\@seccntformat}[1]{\csname the#1\endcsname\ \ }
\makeatother

\newcommand{\N}{\mathbb{N}}
\newcommand{\R}{\mathbb{R}}

\renewcommand{\epsilon}{\ensuremath\varepsilon}

\renewcommand{\phi}{\ensuremath{\varphi}}

\usepackage{mleftright}

\newlist{inlinelist}{enumerate*}{1}
\setlist*[inlinelist,1]{%
  label=(\roman*),
}

\newlist{inlineplainlist}{enumerate*}{1}
\setlist*[inlineplainlist,1]{%
	label=,
}

\DeclarePairedDelimiter\ceil{\lceil}{\rceil}
\DeclarePairedDelimiter\floor{\lfloor}{\rfloor}

\usepackage{makecell}

\usepackage{tikz}
\usetikzlibrary{tikzmark}

\usepackage{xpatch}
\expandafter\xpatchcmd
\csname pgfk@/tikz/every picture/.@cmd\endcsname
{\thepage}{\arabic{page}}{}{}

\tikzstyle{comment} = [draw, fill=blue!70, text=white, text width=3cm, minimum height=1cm, rounded corners, align=left, font=\scriptsize]
\tikzstyle{background_alg} = [draw, fill=blue!20, opacity=0.4, inner sep=4pt, rounded corners=2pt]

\usetikzlibrary{shapes}
\usetikzlibrary{plotmarks}
\usetikzlibrary{calc,fit}

\definecolor{vlgray}{rgb}{0.91 0.91 0.91}
\definecolor{ablack}{rgb}{0.2 0.2 0.2}

\usepackage[customcolors]{hf-tikz}
\hfsetbordercolor{white}
\hfsetfillcolor{vlgray}

\newcounter{highlight}

\newcounter{Ahighlight}

\usepackage{dblfloatfix}

\newcommand{\specialcell}[2][c]{%
 \def\arraystretch{1}\begin{tabular}[#1]{@{}l@{}}#2\end{tabular} \def\arraystretch{1}}

\newcommand{\polylog}[0]{~\text{polylog}}
\newcommand{\polylogshort}[0]{\text{polylog}}

\usepackage{fontawesome}
\usepackage{pifont}
\newcommand{\onestar}{\textsuperscript{*}}

\usepackage{tcolorbox}
\newtcbox{\highlightT}[0]{boxsep=0pt,left=0pt,top=0pt,bottom=0pt,right=0pt,boxrule=0pt,arc=0pt,auto outer arc,colback=green,width=6cm}

\makeatletter
\newcommand{\mybox}[1]{%
  \setbox0=\hbox{#1}%
  \setlength{\@tempdima}{\dimexpr\wd0+13pt}%
  \begin{tcolorbox}[colframe=black,boxrule=0.5pt,arc=4pt,
      left=6pt,right=6pt,top=6pt,bottom=6pt,boxsep=0pt,width=\@tempdima]
    #1
  \end{tcolorbox}
}
\makeatother

\newcommand{\myboxX}[1]{\tikz[baseline={(a.base)}]\node[draw=black,rounded corners=0.5ex,fill=black,inner sep=1pt](a){\textcolor{white}{#1}};}

\newcommand{\noAnswer}{\textcolor{lightgray}{\faQuestionCircle}}

\usepackage{microtype}

\begin{document}




\title{Substream-Centric Maximum Matchings on FPGA}

\author{Maciej Besta}
\affiliation{
  \institution{Department of Computer Science, ETH Zurich}
  \country{Switzerland}
}

\author{Marc Fischer}
\affiliation{
  \institution{Department of Computer Science, ETH Zurich}
  \country{Switzerland}
}

\author{Tal Ben-Nun}
\affiliation{
  \institution{Department of Computer Science, ETH Zurich}
  \country{Switzerland}
}

\author{Dimitri Stanojevic}
\affiliation{
  \institution{Department of Computer Science, ETH Zurich}
  \country{Switzerland}
}

\author{Johannes De Fine Licht}
\affiliation{
  \institution{Department of Computer Science, ETH Zurich}
  \country{Switzerland}
}

\author{Torsten Hoefler}
\affiliation{
  \institution{Department of Computer Science, ETH Zurich}
  \country{Switzerland}
}

\begin{abstract}
Developing high-performance and energy-efficient algorithms for maximum
matchings is becoming increasingly important in social network analysis,
computational sciences, scheduling, and others. In this work, we propose the
first maximum matching algorithm designed for FPGAs; it is
energy-efficient and has provable guarantees on accuracy, performance, and
storage utilization.
To achieve this, we forego popular graph processing paradigms, such as
vertex-centric programming, that often entail large
communication costs. Instead, we propose a \emph{substream-centric} approach,
in which the input stream of data is divided into substreams
processed independently to enable more parallelism while lowering communication
costs.
We base our work on the \emph{theory of streaming graph algorithms} and
analyze 14 models and 28 algorithms. We use this analysis to provide
theoretical underpinning that matches the physical constraints of FPGA
platforms.
Our algorithm delivers high performance (more than 4$\times$ speedup over tuned
parallel CPU variants), low memory, high accuracy, and
effective usage of FPGA resources.
The substream-centric approach could easily be extended to other algorithms to
offer low-power and high-performance graph processing on FPGAs.
\end{abstract}

\ccsdesc[500]{Computer systems organization~Reconfigurable computing}
\ccsdesc[500]{Hardware~Reconfigurable logic and FPGAs}
\ccsdesc[500]{Hardware~Reconfigurable logic applications}
\ccsdesc[500]{Mathematics of computing~Matchings and factors}
\ccsdesc[500]{Mathematics of computing~Graph algorithms}
\ccsdesc[300]{Theory of computation~Parallel algorithms}
\ccsdesc[300]{Mathematics of computing~Approximation algorithms}
\ccsdesc[100]{Theory of computation~Streaming models}
\ccsdesc[100]{Computing methodologies~Vector / streaming algorithms}

\maketitle

\renewcommand{\shortauthors}{M. Besta, M. Fischer, T. Ben-Nun, D. Stanojevic, J. De Fine Licht, and T. Hoefler}

\section{Introduction}
\label{sec:intro}


Analyzing large graphs has become an important task. Example applications
include investigating the structure of Internet links, analyzing relationships
in social media, or capturing the behavior of
proteins~\cite{DBLP:journals/ppl/LumsdaineGHB07, aggarwal2014evolutionary}.
There are various challenges related to the efficient processing of such
graphs. One of the most prominent ones is the size of the graph datasets,
reaching trillions of edges~\cite{ching2015one}. Another one is the fact that
processing such graphs can be very power-hungry~\cite{ahn2016scalable}.

Deriving and approximating \emph{maximum matchings} (MM)~\cite{bondy1976graph}
are important graph problems. A matching in a graph is a set of
edges that have no common vertices. Maximum matchings are used in computational
sciences, image processing, VLSI design, or
scheduling~\cite{trinajstic1986some, bondy1976graph}.  For example, a matching
of the carbon skeleton of an aromatic compound can be used to show the
locations of double bonds in the chemical structure~\cite{trinajstic1986some}.
As deriving the exact MM is usually computationally expensive, significant
focus has been placed on developing fast approximate
solutions~\cite{crouch2014improved}.

To enable high-performance graph processing, various schemes were proposed,
such as vertex-centric approaches~\cite{engelhardt2016vertex},
streaming~\cite{roy2013x}, and others~\cite{simmhan2014goffish}.
They are easily deployable in combination
with the existing processing infrastructure such as
Spark~\cite{zaharia2016apache}. However, they were shown to be often
inefficient~\cite{mcsherry2015scalability} and they are not explicitly
optimized for power-efficiency.

To enable power-efficient graph processing, several graph algorithms and
paradigms for FPGAs were
proposed~\cite{nurvitadhi2014graphgen, engelhardt2016gravf, dai2016fpgp,
zhou2017accelerating, dai2017foregraph, zhang2017boosting,
betkaoui2011framework, zhou2017tunao, weisz2013graphgen, zhou2016high,
oguntebi2016graphops, kapre2015custom}.
Yet, \emph{none targets maximum matchings}. Next, the established
paradigms for designing graph algorithms that were ported to FPGAs, e.g., 
the vertex-centric paradigm, \emph{are not straightforwardly applicable to the
MM problem}~\cite{salihoglu2014optimizing}.

In this work, we propose \emph{the \textbf{first} design and implementation of approximating 
maximum matchings on FPGAs}. Our design is power-efficient \emph{and}
high-performance. For this, we forego the established
vertex-centric paradigm that may result in
complex MM codes~\cite{salihoglu2014optimizing}. Instead,
basing on \emph{streaming theory}~\cite{feigenbaum2005graph},
%
%
we propose a \emph{substream-centric}
FPGA design for deriving MM. In this approach, we \ding{182} divide the incoming stream of
edges into \emph{substreams}, \ding{183} process each substream independently,
and \ding{184} merge these results to form the final algorithm outcome.
%
%
%
%

For highest power-efficiency, phases
\ding{182}--\ding{183} run on the FPGA; both phases work in the
streaming fashion and offer much parallelism, and we identify the FPGA as the
best environment for these phases. Conversely, the final gathering phase, that
usually takes $<1\%$ of the processing time as well as consumed power and
exhibits little parallelism, is conducted on the CPU for more performance.


To provide formal underpinning of our design and thus enable guarantees of
correctness, memory usage, or performance, we base our work
on the family of \emph{streaming models} that were developed to tackle large
graph sizes. A special case is the \emph{semi-streaming
model}~\cite{feigenbaum2005graph}, created specifically for graph processing.
It assumes that the input is a sequence of edges (pairs of vertices), which can
be accessed only sequentially in one direction, as a stream. The main memory
(can be randomly accessed) is assumed to be of size $O(n
\polylog(n))$\footnote{$ O(\polylogshort(n)) = O(\log^c(n))$ for some constant
$c \in \N$} ($n$ is the number of vertices in the graph).  Usually, only
one pass over the input stream is allowed, but some algorithms assume a small
(usually constant or logarithmic) number of passes.
We investigate \emph{a total of 14 streaming models} and \emph{a total of
28 MM algorithms} created in these models, and use the insights from this
investigation to develop our MM FPGA algorithm, ensuring both empirical speedups
and provable guarantees on runtime, used memory, and correctness.

Towards these goals, we contribute:

\begin{itemize}[noitemsep, leftmargin=1em]
\item the first design and implementation of the maximum matching algorithm on FPGAs,
\item an in-depth analysis of the potential of using streaming theory (14 models and 28 algorithms) for accelerating graph processing on FPGAs,
\item a substream-centric paradigm that combines the advantages of semi-streaming theory and FPGA capabilities,
\item detailed {performance analysis demonstrating significant speedups} over state-of-the-art baselines on both CPUs and FPGAs.
\end{itemize}



{This paper is an extended version of our work published in the FPGA'19 proceedings~}\cite{besta2019substream}.

\section{Background and Notation}

We first present the necessary concepts.
\subsection{Graph-Related Concepts}
\label{sec:back_graphs}


\subsubsection{Graph Model}
%
%
We model an undirected graph $G$ as a tuple $(V,E)$; $V = \{v_1, ..., v_n\}$ is
a set of vertices and $E  \subseteq V \times V$ is a set of
edges; $|V|=n$ and $|E|=m$. Vertex labels are $\{1, 2, ..., n\}$.
%
%
%
%
%
If $G$ is weighted, it is modeled by a tuple $(V,E,w)$; $w(e)$ or $w(u,v)$
denote the weight of an edge $e = (u,v) \in E$. The maximum and minimum edge
weight in $G$ are denoted with $w_{max}$ and $w_{min}$.  $G$'s adjacency matrix is
denoted by $A$.
%
%
%
%

\subsubsection{Compressed Sparse Row (CSR)}
In the well-known CSR format, $A$ is represented with three arrays: $val$,
$col$, and $row$. $val$ contains all $A$'s non-zeros (that correspond to $G$'s edges) in the row major
order. $col$ contains the column index for each corresponding value in $val$.
Finally, $row$ contains starting indices in $val$ (and
$col$) of the beginning of each row in $A$. $D$ denotes the diameter of $G$.
%
%
CSR is widely adopted for its simplicity and low memory footprint for sparse
matrices.

\subsubsection{Graph Matching}
\begin{figure}[h!]
\centering
\includegraphics[width=1\textwidth]{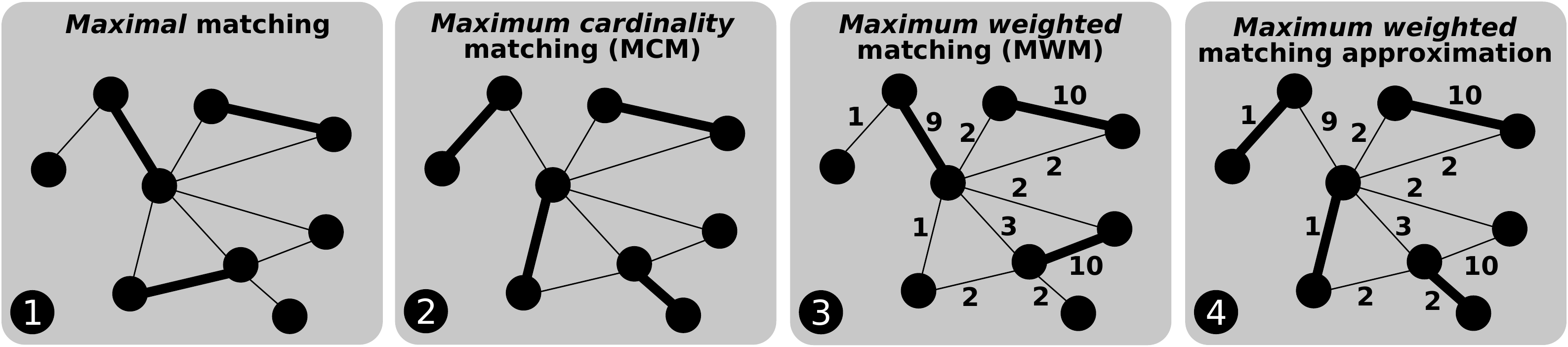}
%
%
%
%
%
\caption{\textbf{Example matchings}. Edges in matchings are represented by bold lines, edge weights are represented with numbers.}\label{fig:algorithm:mwm}
\label{fig:matchings-examples}
\end{figure}

A \emph{matching} $M \subseteq E$ in a graph $G$ is a set of edges that share no vertices.
$M$ is called \emph{maximal} if it is no longer a matching once any edge not in $M$ is
added to it. $M$ is \emph{maximum} if there is no matching with more edges in
it. Maximum matchings (MM) in unweighted graphs are called \emph{maximum
cardinality} matchings (MCM). Maximum matchings in weighted graphs are called
\emph{maximum weighted} matchings (MWM).
Example matchings are illustrated in Figure~\ref{fig:matchings-examples}.


\subsubsection{Maximum Weighted Matching}
Given a weighted graph $G=(V, E, w)$, a maximum weighted matching is a matching
$M^*$, such that its weight $w(M^*) = \sum_{e \in M^*} w(e)$ is maximized.
%
%
{
An algorithm
provides an $\alpha$-approximation of $M^*$, if -- for any derived weighted matching $M$ -- it holds that
$w(M^*) / w(M) \leq \alpha$ (therefore $\alpha \geq 1$).
Note that the approximation ratio of the MCM is defined inversely compared to
the MWM: We say that an algorithm, that returns a matching $M_C$,
provides an $\alpha$-approximation to the maximum cardinality $M^*_C$
if it holds that $|M_C|/|M^*_C| \geq \alpha$ (therefore $\alpha \leq 1$).
We do this to conform to the general approximation notation of
maximum cardinality matching }\cite{karp1990optimal, kapralov2013better,
konrad2012maximum}{ and maximum weighted matching }\cite{ghaffari2017space,
grigorescu2016streaming, crouch2014improved}.

\subsection{Architecture-Related Concepts}

%


\subsubsection{FPGAs}
FPGAs aim to combine the advantages of
Application Specific Integrated Circuits (ASICs) and CPUs: they offer ASIC's
high performance and low power usage, and they can be reconfigured to {compute arbitrary
functions, similarly to CPUs.}
Usually, the FPGA clock frequency is $\approx${10--600}MHz, dependent on the
algorithm and the FPGA platform. This is an order of magnitude less compared to
high-end CPUs (up to 4.7GHz~\cite{intelcpu}) and below GPUs (up to
1.5GHz~\cite{nvidiagpu}). Yet, due to the custom design deployed directly in
hardware, multiple advantages such as low power consumption arise~\cite{de2018transformations}.


\subsubsection{FPGA Components {and Fundamental Building Blocks}}
{Xilinx uses vanilla look-up tables (LUTs) while Intel employs Adaptive Logic
Modules as fundamental building blocks~}\cite{tyhach2015arria, shang2002dynamic}{. Their micro-architecture is different
but fundamental functionality is similar. Xilinx combines multiple LUTs and
associated registers into CLBs~}\cite{shang2002dynamic}{, while Intel combines ALMs into LABs~}\cite{tyhach2015arria}.
%
%
%
%
Next, Block Random Access Memory (BRAM) allows to store small amounts of data
(up to 20 kbits per BRAM~\cite{intelfpga}) and provides fast data access, acting
similarly to a CPU cache. {Today, thousands} of BRAM units are
distributed over a single FPGA. 

\subsubsection{FPGA+CPU}
Hybrid computation systems consist of a host CPU and an attached FPGA. First
\ding{182}, an FPGA can be added to the system as an accelerator; the host main
memory is separated from the FPGA private DRAM memory and data must be
transferred over PCIe. Often, the FPGA is configured as a PCIe endpoint with a
direct memory access (DMA) controller, allowing to move data between the host
and the FPGA without the need of CPU resources. PCIe is high-bandwidth
oriented, but exhibits high overhead and latency for small
packets~\cite{choi2016quantitative}. This drawback is overcome by storing often
accessed data in the private DRAM using the memory controller, or storing the
data on chip in the FPGA's BRAM.
Second \ding{183}, the CPU and the FPGA can be directly linked by an
interconnect, such as Intel's QuickPath Interconnect (QPI), providing a
coherent view to a single shared main memory. Examples of these systems include
Intel HARP~\cite{oliver2011reconfigurable} and the Xilinx Extensible Processing
Platform~\cite{santarini2011zynq}. The direct main memory access allows to
share data without the need to copy it to the FPGA. To prevent direct physical
main memory accesses, HARP provides a translation layer, allowing the FPGA to
operate on virtual addresses. It is implemented in both hardware as a System
Protocol Layer (SPL) and in software, for example as a part of the Centaur
framework~\cite{owaida2017centaur}. Moreover, a cache is available to reduce
access time.  According to Choi et al.~\cite{choi2016quantitative}, systems
with direct interconnect exhibit lower latency and higher throughput than PCIe
connected FPGAs.
\emph{In our substream-centric FPGA design for deriving MM, we use a hybrid
CPU+FPGA system to take advantage of both the CPU and the FPGA in the context
of graph processing}.

%

\section{From Semi-Streaming to FPGAs}
\label{sec:theory-analysis}

We first summarize the analysis into the theory of streaming models and
algorithms. We conducted the analysis to provide formal underpinning of our
work and thus \emph{\textbf{ensure provable properties, for example correctness,
approximation, or performance}}. Towards this goal, \emph{we analyzed 14
different models of streaming} (simple streaming~\cite{henzinger1998computing},
semi-streaming~\cite{feigenbaum2005graph},
insert-only~\cite{feigenbaum2005graph}, dynamic~\cite{ahn2012graph},
vertex-arrival~\cite{cormode2018independent},
adjacency-list~\cite{mcgregor2016better},
cash-register~\cite{muthukrishnan2005data},
Turnstile~\cite{muthukrishnan2005data}, sliding
window~\cite{datar2002maintaining}, annotated
streaming~\cite{chakrabarti2009annotations},
StreamSort~\cite{aggarwal2004streaming}, W-Stream~\cite{demetrescu2009trading},
online~\cite{karande2011online}, and MapReduce~\cite{dean2008mapreduce}) and
\emph{28 different MM algorithms}. 
%
%
Moreover, to understand whether streaming itself is the best option for
developing MWM on FPGAs, we extensively analyzed existing graph processing
works on FPGAs~\cite{khoram2018accelerating, yao2018efficient, zhang2018degree,
yang2018efficient, zhou2018framework, lee2017extrav, ma2017fpga,
zhou2017accelerating, zhang:graph_FPGA, dai2017foregraph, lei2016fpga,
ozdal2016energy, zhou2016high, engelhardt2016gravf, oguntebi2016graphops,
dai2016fpgp}.
{The outcome of this analysis is as follows: the best candidates for adoption
in the FPGA setting are \textbf{semi-streaming} graph algorithms that
\textbf{expose parallelism by decomposing the incoming stream of edges for
independent processing}, for example the MWM algorithm by Crouch and
Stubbs~\cite{crouch2014improved}}.

\subsection{Why Streaming (aka Edge-Centric)?}

Before analyzing different streaming models and algorithms, we first
investigate whether streaming itself is the best paradigm for developing the
MWM algorithm for FPGAs. In the streaming paradigm, also
known as the \emph{edge-centric} paradigm, the ``first class citizen'' is an
edge. A sequence of edges is streamed in and out from the memory to the FPGA,
and a specified operation is performed on each edge and possibly some
associated edge properties.  This way of accessing the graph has major
advantages because of its sequential memory access pattern, which improves
spatial locality~\cite{besta2019graph}.


However, there also exist other paradigms for processing graphs, most
importantly the established vertex-centric paradigm~\cite{malewicz2010pregel},
where the ``first class citizen'' is a vertex.  Here, one programs an algorithm
by developing a (usually small) routine that is executed \emph{for each vertex
in the graph concurrently}. In this routine, one usually has access to the
neighbors of a given vertex.
Such an approach can lead to many random memory accesses as neighboring
vertices may be stored in different regions of the memory. Still, it is often
used because many important algorithms such as BFS or PageRank can easily be
implemented in this model~\cite{besta2019graph}.

To identify the best graph processing paradigm for implementing the MWM
algorithm in FPGA, we first analyze the existing FPGA graph processing
implementations, focusing on the used paradigm. Table~\ref{tab:papers}
illustrates the most relevant designs. We group separately generic graph
processing frameworks and specific algorithm implementations. Each group is
sorted chronologically. Selected columns in this table constitute criteria used
to categorize the surveyed FPGA works (the full results of this analysis are in
a separate extended survey~\cite{besta2019graph}).

The first such criterion is \textbf{generality}, i.e., whether a given FPGA
scheme is focused on a particular graph problem or whether it constitutes a
generic framework that facilitates implementing different graph algorithms.
Another criterion is a used \textbf{graph programming paradigm}. 
%
%
%
We also distinguish between works that target a single FPGA and ones that scale
to \textbf{multiple FPGAs}. Finally, we consider the used \textbf{programming
language} and the \textbf{storage location} of the \emph{whole} processed graph
datasets. In the latter, ``DRAM'' indicates that the
input dataset is located in DRAM and it is streamed in and out
of the FPGA during processing (i.e., only a part of the input dataset is stored
in BRAM at a time). Contrarily, ``BRAM'' indicates that the whole dataset is
assumed to be located in BRAM.  To investigate the scalability of the analyzed
solutions, we provide sizes ($n, m$) of the largest processed graphs.

The analysis indicates that the \emph{streaming (edge-centric) paradigm and its
variants have so far been the most successful in processing large graphs}. The
only vertex-centric design that processed a graph with $m >$1B required
multiple FPGAs~\cite{betkaoui2012reconfigurable}. Contrarily, two recent
edge-centric designs based on single FPGAs were able to conduct computations on
such graphs~\cite{lee2017extrav, zhou2018framework}.

Moreover, although the vertex-centric paradigm facilitates developing
simple algorithms such as BFS or PageRank, it is often difficult to use for
solving more complex graph problems.
For example, as Salihoglu and Widom state~\cite{salihoglu2014optimizing},
\emph{``(...) implementing graph algorithms efficiently on Pregel-like systems
(...) can be surprisingly difficult and require careful optimizations.''}.  For
example, when describing graph problems as fundamental as deriving Strongly
Connected Components (SCCs), Bi-Connected Components (BCCs), and Minimum
Spanning Forest (MSF), Salihoglu and Widom~\cite{salihoglu2014optimizing}
observe that \emph{``(...) implementing even the basic versions of SCC and MSF
is quite challenging, taking more than 700 lines of code.''} while Yan et
al.~\cite{yan2014pregel} state that \emph{``It is challenging to design Pregel
algorithms for problems such as BCCs and SCCs.''}.
The problem is not only related to Pregel. Yan et al.~\cite{yan2014pregel} make
similar observations about the established GraphLab~\cite{low2010graphlab} and
PowerGraph~\cite{gonzalez2012powergraph} vertex-centric frameworks, stating
that they do \emph{``not support algorithms in which a vertex needs to
communicate with a non-neighbor, such as the }[Shiloach-Vishkin]\emph{ algorithm }[for
Connected Components]\emph{, the list ranking algorithm, and the }[Case Checking]
\emph{algorithm.}''.  They make similar observations for BCCs and SCCs.
Thus, when developing vertex-centric graph algorithms, one resorts to
algorithms that fit the vertex-centric paradigm well. An example is Label
Propagation for Connected Components~\cite{yan2014pregel}. Yet, this algorithm
takes $O(D)$ time in the PRAM model~\cite{harris1994survey} while
the Shiloach-Vishkin algorithm~\cite{shiloach1980log}, hard to express in the
vertex-centric paradigm~\cite{yan2014pregel}, uses only $O(\log n)$ time in PRAM. Similar observations are made for other
graph problems~\cite{khan2016vertex}.
This indicates that \emph{it is difficult to design efficient vertex-centric
formulations of graph algorithms that require accessing more than neighbors of
each vertex}.

\emph{\textbf{Thus, in our work, we use edge streaming for developing the MWM
algorithm for FPGAs.}} This is because (1) it has been shown to scale to large
graphs and (2) it straightforwardly enables pipelining of edges, thus
facilitating the utilization of FPGA hardware resources. Finally (3), the
existing rich theory of streaming graph processing for complex graph problems
such as matchings, random walks, and others~\cite{feigenbaum2005graph,
feigenbaum2005graph, ahn2012graph, cormode2018independent, mcgregor2016better,
muthukrishnan2005data, muthukrishnan2005data, datar2002maintaining,
chakrabarti2009annotations, aggarwal2004streaming, demetrescu2009trading,
karande2011online, dean2008mapreduce} indicates that it is easier to
develop fast MWM schemes with edge streaming than with the vertex-centric paradigm.

\begin{table}[hbtp]
\centering
 \setlength{\tabcolsep}{0.6pt}
\def\arraystretch{1.4}
\footnotesize
\begin{tabular}{@{}llllllllrr@{}}
\toprule
%
  \makecell[c]{\textbf{Reference}\\\textbf{(scheme name)}} & \makecell[c]{\textbf{Venue}} & \makecell[c]{\textbf{Generic}\\\textbf{Design}$^1$} & \makecell[c]{\textbf{Considered}\\\textbf{Problems$^2$}} & \makecell[c]{\textbf{Programming}\\\textbf{Paradigm$^4$}}
  & \makecell[c]{\textbf{Used}\\\textbf{Language}}
  & \makecell[c]{\textbf{Multi}\\\textbf{FPGAs}$^4$}
  & \makecell[c]{\textbf{Input} \\\textbf{Location}$^5$} & $n^{\text{\textdagger}}$ & $m^{\text{\textdagger}}$ \\
%
\midrule
\makecell[l]{Kapre \cite{kapre2006graphstep}\\\textbf{(GraphStep)}}             & FCCM'06       & \faThumbsOUp & \makecell[l]{spreading\\activation~\cite{liu2004conceptnet}}   & \makecell[l]{BSP (similar to\\vertex-centric)}     &unsp. & \faThumbsOUp & BRAM       & 220k   &  550k    \\ 
\makecell[l]{Weisz \cite{weisz2013graphgen}\\\textbf{(GraphGen)}}                     & FCCM'14       & \faThumbsOUp & \makecell[l]{TRW-S,\\CNN~\cite{sun2003stereo}}  & Vertex-Centric   &unsp. &\faThumbsDown & DRAM  & 110k    & 221k  \\ 
\makecell[l]{Kapre \cite{kapre2015custom}\\\textbf{(GraphSoC)}}            & ASAP'15       & \faThumbsOUp & SpMV                        & Vertex-Centric  & \makecell[l]{C++\\(HLS)}  & \faThumbsOUp & BRAM  & 17k     & 126k  \\ 
\makecell[l]{Dai \cite{dai2016fpgp}\\\textbf{(FPGP)}}                             & FPGA'16       & \faThumbsOUp  & BFS                         & \makecell[l]{Edge-Centric$^*$}                      &unsp. & \faThumbsOUp & DRAM  & 41.6M   & \textbf{1.4B} \\ 
\makecell[l]{Oguntebi \cite{oguntebi2016graphops}\\\textbf{(GraphOps)}}               & FPGA'16       & \faThumbsOUp & \makecell[l]{BFS, SpMV, PR,\\Vertex Cover}      & Edge-Centric$^*$                      & \makecell[l]{MaxJ\\(HLS)}  &\faThumbsDown & BRAM  & 16M     & 128M  \\ 
Zhou \cite{zhou2016high}         & FCCM'16       & \faThumbsOUp & \makecell[l]{SSSP, WCC, MST}                & Edge-Centric          &unsp. &\faThumbsDown & DRAM  & 4.7M    & 65.8M \\ 
\makecell[l]{Engelhardt \cite{engelhardt2016gravf}\\\textbf{(GraVF)}}           & FPL'16        & \faThumbsOUp & \makecell[l]{BFS, PR, SSSP, CC}              & Vertex-Centric        & \makecell[l]{Migen\\(HLS)} &\faThumbsDown & BRAM  & 128k    & 512k \\ 
\makecell[l]{Dai \cite{dai2017foregraph}\\\textbf{(ForeGraph)}}                        & FPGA'17       & \faThumbsOUp & \makecell[l]{PR, BFS, WCC}                &  Edge-Centric$^*$                     &unsp. & \faThumbsOUp & DRAM  & 41.6M  & \textbf{1.4B}  \\ 
Zhou \cite{zhou2017accelerating} & SBAC-PAD'17   & \faThumbsOUp &  BFS, SSSP            & \makecell[l]{Hybrid (Vertex\\+Edge-Centric)}  &unsp. &\faThumbsDown & DRAM  & 10M    & 160M   \\ 
\makecell[l]{Lee \cite{lee2017extrav}\\\textbf{(ExtraV)}}                        & FPGA'17       & \faThumbsOUp & \makecell[l]{BFS, PR, CC} & Edge-Centric$^*$  & \makecell[l]{C++\\(HLS)}  &\faThumbsDown & DRAM  & 124M   & \textbf{1.8B}  \\ 
Zhou \cite{zhou2018framework}     & CF'18         & \faThumbsOUp & SpMV, PR                    & Edge-Centric    &unsp.      &\faThumbsDown & DRAM  & 41.6M  & \textbf{1.4B}  \\ 
Yang \cite{yang2018efficient}                   & report (2018)        & \faThumbsOUp & BFS, PR, WCC                & Edge-Centric$^*$                      & OpenCL  &\faThumbsDown &   & 4.85M  & 69M   \\ 
Yao \cite{yao2018efficient}                     & report (2018)        & \faThumbsOUp & BFS, PR, WCC                &      Vertex-Centric                 &unsp.      &\faThumbsDown & BRAM  & 4.85M  & 69M  \\ 
\midrule
Betkaoui \cite{betkaoui2011framework}           & FTP'11        & \faThumbsDown & GC                                    & Vertex-Centric        & Verilog  & \faThumbsOUp & DRAM  & 300k  & 3M     \\
Betkaoui \cite{betkaoui:APSP_FPGA}              & FPL'12        & \faThumbsDown & APSP                                  & Vertex-Centric        & Verilog  & \faThumbsOUp & DRAM  & 38k     & 72M    \\ 
Betkaoui \cite{betkaoui2012reconfigurable}       & ASAP'12       & \faThumbsDown & BFS                                   & Vertex-Centric        & Verilog  & \faThumbsOUp & DRAM  & 16.8M  & 1.1B    \\ 
\makecell[l]{Attia~\cite{attia2014cygraph}\\\textbf{(CyGraph)}}                   & IPDPS'14      & \faThumbsDown & BFS                                   & Vertex-Centric        & VHDL  & \faThumbsOUp & DRAM  & 8.4M    & 536M  \\ 
Zhou \cite{zhou2015pagerank}      & ReConFig'15   & \faThumbsDown & PR                                    & Edge-Centric          &unsp. &\faThumbsDown & DRAM  & 2.4M    & 5M    \\ 

%
Besta \cite{besta2019substream} & FPGA'19 & \faThumbsDown & MWM & Substream-Centric & Verilog & \faThumbsDown & DRAM & 4.8M & 117M \\
\bottomrule
\end{tabular}
\caption{
Summary of the features of selected works sorted by publication date.
$^1$\textbf{Generic Design}: this criterion indicates whether a given scheme
provides a graph processing framework that supports more than one graph algorithm (\faThumbsOUp)
or whether it focuses on concrete graph algorithm(s) (\faThumbsDown).
$^2$\textbf{Considered Problems}: this column lists graph problems (or
algorithms) that are explicitly considered in a given work; they are all
described in detail in an extended survey~\cite{besta2019graph} (BFS:
Breadth-First Search~\cite{Cormen:2001:IA:580470}, SSSP: Single-Source Shortest
Paths~\cite{Cormen:2001:IA:580470}, APSP: All-Pairs Shortest
Paths~\cite{Cormen:2001:IA:580470}, PR: PageRank~\cite{page1999pagerank}, CC:
Connected Components~\cite{Cormen:2001:IA:580470}, WCC: Weakly Connected
Components~\cite{Cormen:2001:IA:580470}, MST: Minimum Spanning
Tree~\cite{Cormen:2001:IA:580470}, SpMV: Sparse Matrix and Dense Vector
product, TC: Triangle Counting~\cite{schank2007algorithmic}, BC: Betweenness
Centrality~\cite{newman2005measure}, GC: Graphlet
Counting~\cite{schank2007algorithmic}, TRW-S: Tree-Reweighted Message
Passing~\cite{szeliski2008comparative}, CNN: Convolutional Neural
Networks~\cite{cambriconx, ben2019modular}).
$^3$\textbf{Used Programming Paradigm}: this column specifies programming paradigms and models used
in each work;
$^*$The star indicates that a given scheme uses a paradigm similar to the edge-centric
streaming paradigm, for example sharding as used in GraphChi~\cite{kyrola2012graphchi}, where edges are
first preprocessed and divided into shards, with shards being streamed in and out of the main memory.
$^4$\textbf{Multi FPGAs}: this criterion indicates whether a given scheme scales to
multiple FPGAs (\faThumbsOUp) or not (\faThumbsDown).
$^5$\textbf{Input Location}: this column indicates the location of the \emph{whole} input graph dataset.
``DRAM'' indicates that it is located in DRAM
and it is streamed in and out of the FPGA during processing (i.e., only a part of
the input dataset is stored in BRAM at a time). Contrarily, ``BRAM''
indicates that the whole dataset is assumed to be located in BRAM.
$n^\text{\textdagger}, m^\text{\textdagger}$: these two columns contain the numbers of
vertices and edges used in the largest graphs considered in respective works.
%
%
In any of the columns, ``unsp.'' indicates that a given value is not specified.
}
\label{tab:papers}
\end{table}

\subsection{Why Semi-Streaming?}

The \emph{semi-streaming model}~\cite{feigenbaum2005graph} was created
specifically for graph processing. 
However, there are numerous other streaming models that are also used for
developing graph algorithms, namely simple
streaming~\cite{henzinger1998computing},
insert-only~\cite{feigenbaum2005graph}, dynamic~\cite{ahn2012graph},
vertex-arrival~\cite{cormode2018independent},
adjacency-list~\cite{mcgregor2016better},
cash-register~\cite{muthukrishnan2005data},
Turnstile~\cite{muthukrishnan2005data}, sliding
window~\cite{datar2002maintaining}, annotated
streaming~\cite{chakrabarti2009annotations},
StreamSort~\cite{aggarwal2004streaming}, W-Stream~\cite{demetrescu2009trading},
online~\cite{karande2011online}, and MapReduce~\cite{dean2008mapreduce}.  A
detailed comparison of these models and analysis of their relationships is
outside the scope of this paper\footnote{This comparison will be provided in a
separate survey on streaming graph processing; this survey will be released
upon the publication of this work.}.  Here, we briefly justify why we selected
semi-streaming as the basis for our MWM algorithm for FPGAs.  First,
semi-streaming enables a generic streaming setting in which one can arbitrarily
process the incoming stream of edges.  This will enable our substream-centric
approach where the incoming edges are divided into independent substreams.
Second, there exists a rich body of algorithms and techniques developed for the
semi-streaming setting.  Finally, most importantly, in semi-streaming one
assumes that processing the incoming stream of edges can utilize at most $O(n
\polylog(n))$ random memory.  \emph{\textbf{Thus, algorithms under this model
may address the limited FPGA BRAM capacity better than algorithms in models
with weaker memory-related constraints}}.


\subsection{Which Semi-Streaming MM Algorithm?}
\label{sec:algs-theory}


Table~\ref{table:mcm} compares the considered
semi-streaming and related MM algorithms.
We identify those with
properties suggesting an effective and versatile FPGA design:
low space consumption, one pass, and applicability
to general graphs.
Finally, virtually all designed algorithms are approximate.
Yet, as we show later (\cref{sec:evaluation}), in practice they
deliver near-accurate results.

\begin{table}[hbtp]
%
\setlength{\tabcolsep}{1.5pt}
%
\begin{minipage}{\columnwidth}
\begin{center}
\def\arraystretch{1.4}
\begin{tabular}{lllllll}
\toprule
\textbf{Reference} & \textbf{Approx.} & \textbf{Space} & \textbf{\#Passes} & \textbf{Wgh}$^1$ & \textbf{Gen}$^2$ & \textbf{Par}$^3$ \\ \midrule
\cite{feigenbaum2005graph} & $1/2$ & $O(n)$ & $1$ & \faThumbsDown & \faThumbsOUp & \faThumbsUp \\ 
\cite[Theorem 6]{konrad2012maximum} & $1/2 + 0.0071$ & $O( n \polylog(n))$ & 2 & \faThumbsDown & \faThumbsOUp & \faThumbsUp \\ 
\cite[Theorem 2]{konrad2012maximum} & $1/2 + 0.003$\onestar & $O(n \polylog(n))$ & $1$ & \faThumbsDown & \faThumbsOUp & \faThumbsUp  \\ 
\cite[Theorem 1.1]{kapralov2014approximating} & $O(\polylogshort(n))$ & $ O(\polylogshort(n))$ & $1$ & \faThumbsDown & \faThumbsOUp & \faThumbsOUp \\ 
\cite[Theorem 1]{feigenbaum2005graph} & $2/3-\epsilon$ & $O(n \log n)$ & $O\left( \log \left( 1 / \epsilon \right)  / \epsilon\right)$ & \faThumbsDown & \faThumbsDown & \faThumbsUp \\ 
\cite[Theorem 19]{ahn2011linear} & $1- \epsilon$ & $ O\left(n \polylog(n) / \epsilon^2 \right) $ & $O\left(\log \log \left(1 / \epsilon\right) / \epsilon^2 \right) $ & \faThumbsDown & \faThumbsDown & \faThumbsUp \\ 
\cite[Theorem 5]{konrad2012maximum} & $1/2 + 0.019$ & $O(n \polylog(n))$ & $2$ & \faThumbsDown & \faThumbsDown & \faThumbsUp \\ 
\cite[Theorem 1]{konrad2012maximum} & $1/2 + 0.005$\onestar & $O(n \log n)$ & $1$ & \faThumbsDown & \faThumbsDown & \faThumbsUp \\ 
\cite[Theorem 4]{konrad2012maximum} & $1/2 + 0.0071$\onestar& $O(n \polylog(n))$ & $2$ & \faThumbsDown & \faThumbsDown & \faThumbsUp \\ 
\cite{karp1990optimal} & $1-1/e$ & $O(n \polylog(n))$ & 1 & \faThumbsDown & \faThumbsDown  & \faThumbsUp \\ 
\cite[Theorem 20]{goel2012communication}  & $1-1/e$ & $O(n)$ & $1$ & \faThumbsDown & \faThumbsDown & \faThumbsUp \\ 
\cite[Theorem 2]{kapralov2013better} & $ 1 - \frac{e^{-k } k ^{k-1}}{(k-1)!}$ & $O(n)$ & $k$ & \faThumbsDown & \faThumbsDown & \faThumbsUp \\ 
\cite{chitnis2016kernelization} & 1 & $\tilde{O}\left( k ^2 \right) $ & $1$ & \faThumbsDown & \faThumbsOUp & \faThumbsOUp \\ 
\cite{chitnis2016kernelization} & $1/\epsilon$ & $\tilde{O}\left( n^2 / \epsilon^{3}\right)$ & 1 & \faThumbsDown & \faThumbsOUp & \faThumbsOUp \\ 
\cite[Theorem 1]{assadi2016maximum} & $1/n^\epsilon$ & $\tilde{O}\left(n^{2-3\epsilon} + n^{1-\epsilon}\right)$ & 1 & \faThumbsDown & \faThumbsDown & \faThumbsOUp \\
\midrule
\cite[Theorem 2]{feigenbaum2005graph} & $6$ & $O(n \log n)$ & 1 & \faThumbsOUp & \faThumbsOUp & \noAnswer \\  
\cite[Theorem 3]{mcgregor2005finding} & $2+\epsilon$ & $O( n \polylog(n))$ & $O(1)$ & \faThumbsOUp & \faThumbsOUp & \noAnswer \\ 
\cite[Theorem 3]{mcgregor2005finding} & $5.82$ & $O( n \polylog(n))$ & $1$ & \faThumbsOUp & \faThumbsOUp & \noAnswer \\ 
\cite{zelke2012weighted} & $5.58$ & $O(n \polylog(n))$ & 1 & \faThumbsOUp & \faThumbsOUp & \noAnswer \\ 
\cite{epstein2011improved} & $4.911+\epsilon$ & $O( n \polylog(n))$ & 1 & \faThumbsOUp & \faThumbsOUp & \noAnswer \\ 
\cite{grigorescu2016streaming} & $3.5+\epsilon$ & $O(n \polylog(n))$ & 1 & \faThumbsOUp & \faThumbsOUp & \noAnswer \\ 
\cite{paz20172+} & $2+\epsilon$ & $O\left(n \log^2 n \right)$ & 1 & \faThumbsOUp & \faThumbsOUp & \noAnswer \\ 
\cite{ghaffari2017space} & $2 + \epsilon$ & $O(n \log n)$ & 1 & \faThumbsOUp & \faThumbsOUp & \noAnswer \\ 
\cite[Section 3.2]{feigenbaum2005graph} & $ 2+\epsilon$ & $ O(n \log n)$ & $O\left( \log_{1+\epsilon/3} n\right)$ & \faThumbsOUp & \faThumbsOUp & \noAnswer \\ 
\cite[Theorem 28]{ahn2011linear} & $\frac{1}{1-\epsilon}$ & $O\left( n \log (n) / \epsilon^4 \right)$ & $O\left( \epsilon^{-4}  \log n  \right) $ & \faThumbsOUp & \faThumbsOUp & \faThumbsUp \\ 
\cite[Theorem 22]{ahn2011linear} & $\frac{1}{\frac{2}{3}(1-\epsilon)}$ & $O\left( n \left(\frac{\epsilon \log n - \log \epsilon}{\epsilon^2}\right) \right) $ & $O\left(\epsilon^{-2} \log\left(\epsilon^{-1}\right)\right)$ & \faThumbsOUp & \faThumbsOUp & \faThumbsUp \\ 
\cite[Theorem 22]{ahn2011linear} & $ \frac{1}{1-\epsilon}$ & $O\left( n \left(\frac{\epsilon \log n - \log \epsilon}{\epsilon^2}\right) \right) $  & $O\left(\epsilon^{-2} \log\left(\epsilon^{-1}\right)\right)$ & \faThumbsOUp &\faThumbsDown & \faThumbsUp \\
\midrule
\textbf{\cite{crouch2014improved}} & $4+\epsilon$ & $O(n \polylog(n))$ & 1 & \faThumbsOUp & \faThumbsOUp & \faThumbsOUp \\ 
  \bottomrule
\end{tabular}
\end{center}
\end{minipage}
\caption{(\cref{sec:theory-analysis}) \textbf{Comparison of algorithms for maximum matching}. 
%
%
$^*$Approximation in expectation, 
$^1$\textbf{Wgh}: accepted weighted graphs,
$^2$\textbf{Gen}: accepted general (non-bipartite) graphs,
$^3$\textbf{Par}: Potential for parallelization;
$k$ is the size of a given maximum matching.
\faThumbsOUp: A given feature is offered. 
\faThumbsDown: A given
feature is not offered.
In the context of parallelization: 
\faThumbsOUp: a given algorithm is based on a method
that is easily parallelizable (e.g., sampling),
\faThumbsUp: a given algorithm uses a method that may be complex
to parallelize (e.g., augmenting paths),
\noAnswer: it is unclear how to parallelize a given
algorithm (e.g., it is based on a greedy approach). 
}
\label{table:mcm}
\end{table}

We conjecture that the majority of the considered MM algorithms deliver limited performance on
FPGA because \emph{their design is strictly sequential}: every edge in the
incoming stream can only be processed after processing the previous edge in the
stream is completed.
However, we identify some algorithms that introduce a certain amount of
parallelism. Here, we focus on the algorithm by Crouch and
Stubbs~\cite{crouch2014improved}, used as a basis for our
FPGA design (last row of Table~\ref{table:mcm}). We first outline this algorithm and then
justify our selection.
We also discuss other considered MWM algorithms.

\subsubsection{Algorithm Intuition}
\begin{figure}[h!]
\centering
\includegraphics[width=1\textwidth]{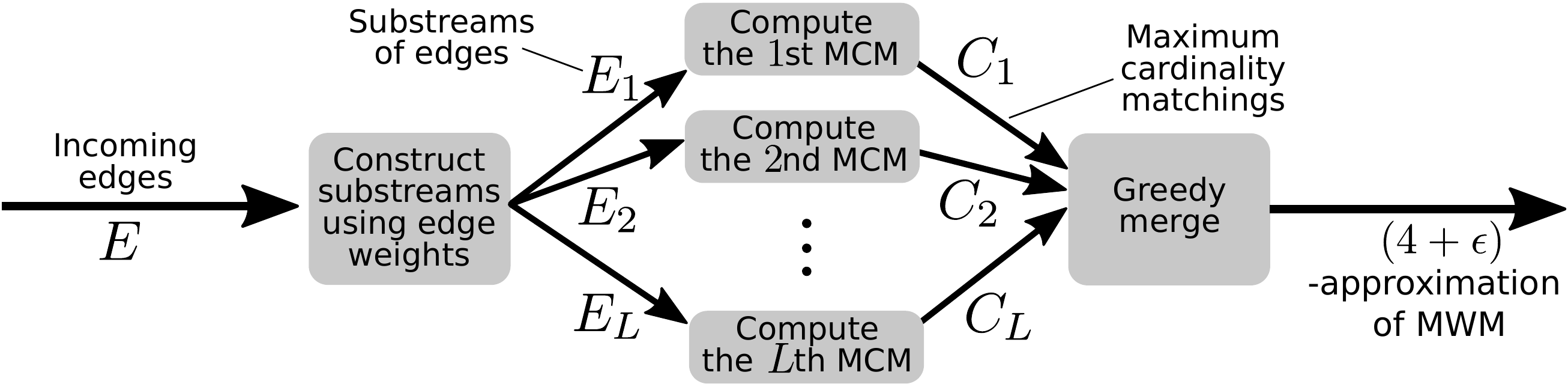}
%
\caption{\textbf{The design of the MWM algorithm}
of Crouch and Stubbs~\cite{crouch2014improved}.}
%
\label{fig:algorithm:mwm:block}
\end{figure}

The MWM algorithm by Crouch and Stubbs~\cite{crouch2014improved}
delivers a $(4+\epsilon)$-approximation of MWM. It consists of
two parts.  In Part~1, one selects $L$ subsets of the incoming (streamed) edges and
computes a \emph{maximum cardinality} matching for each such subset.  In
Part~2, the derived maximum matchings are combined into the final \emph{maximum
weighted matching}.
The approach is visualized in Figure~\ref{fig:algorithm:mwm:block}. 

\subsubsection{Algorithm Details}
The algorithm of Crouch and Stubbs~\cite{crouch2014improved} provides a
$(4+\epsilon)$-approximation to the MWM problem assuming an {arbitrarily ordered} stream of
incoming edges with possible graph updates (edge insertions).  The basic idea
is to reduce the MWM problem to $L \equiv O(\polylogshort(n))$ instances of the MCM
problem. Given the input stream of incoming edges $E$, $O\left(\frac{1}{\epsilon}
\log n\right)$ many substreams are generated. Each substream $E_i$ is created by
filtering the edges according to their weight. Specifically, we have $E_i = \{
e \in E\ |\ w(e) \geq ( 1 + \epsilon)^i \}$.  Since an edge that belongs to
substream~$i+1$ also belongs to substream~$i$, it holds that $E_{i+1} \subseteq
E_i$.  
Next, for each substream, an MCM $C_i$ is constructed. The final $(4+
\epsilon)$-approximation to MWM is greedily constructed by considering the
edges of every $C_i$, in the descending order of~$i$. 

We select this algorithm as the basis of our substream-centric FPGA design as
it (1) can be straightforwardly parallelized, (2) ensures only $O(n \polylog
n)$ memory footprint as it belongs to the semi-streaming model, (3) targets
general weighted graphs, (4) its structure matches well the design of a hybrid
FPGA+CPU system: while substreams can be processed in parallel on the FPGA, the
greedy sequential merging of substreams into the final MWM can be done on the
CPU, and (5) it needs only one pass over the streamed dataset of size $O(m+n)$,
limiting expensive data transfers between the FPGA and DRAM.

\subsection{How To Adapt Semi-Streaming to FPGAs?}

In~\cref{sec:main}, we describe the FPGA adaptation, design, and implementation
of the selected semi-streaming MM algorithm.  We stream the edges as stored in
the CSR representation. Our substream-centric design implements a staged
pipeline with throughput of up to one edge per cycle.

\subsection{Other Considered Matching Algorithms}
\label{sec:other-algs}

We could not find other algorithms that would clearly satisfy all the above
five criteria (stated in~\cref{sec:algs-theory}) simultaneously. However, we do
not conclude that other algorithms are unsuitable for an efficient FPGA
implementation, and leave developing such designs as future work. Here, for
completeness, we describe other streaming matching algorithms considered in our
analysis. Details of these algorithms (models, techniques, random vs.~deterministic
design) can be found in Table~\ref{tab:algs-2}.

\subsubsection{Generic Techniques for Deriving Matchings}
\label{sec:general-techniques}
First, we shortly describe generic techniques
that we identified while analyzing matching algorithms.
We identified these techniques to investigate which algorithms
are easily parallelizable (when a given technique is easily parallelizable,
this implies that algorithms relying on this technique may also be easily
parallelizable).

\noindent
\macb{Sampling and Unbiased Estimators} 
Sampling~\cite{dixon1957introduction} in general is used to estimate some
quantity, for example the number of triangles in a graph.
%
%
One first samples edges that are then used to generate an unbiased estimator of
the graph property in question.  After that, the challenging task is to show
that the space constraint of $O( n \polylog(n))$ is not exceeded, and that the
estimator succeeds to give an estimation within a small error range with some
probability. Often, the error range and success probability can be controlled
and are assumed to be constants.  This method is abbreviated with a simple
\emph{General sampling}.

\noindent
\macb{Sketching and $L_0$ Sampling}
Graph sketching is a technique that reduces dimensionality of given data while
preserving certain properties. It is commonly used in dynamic graph streams
because in such streams a sampled edge might be removed afterwards, and thus a
simple sampling scheme as described above cannot be used.
Specifically, sketching based on \emph{$L_0$ sampling} proved to be useful
in different cases~\cite{ahn2012analyzing, ahn2012graph} when the dynamic
streaming model with inserts and deletes is assumed. 


\noindent
\macb{Simulation of Local Distributed Algorithms}
An approach used in several cases is
based on porting a local distributed algorithm solving a given graph problem to
the streaming model to simulate its behaviour and solve the same problem in the
streaming setting~\cite{kapralov2014approximating}.

\noindent
\macb{Augmenting Paths}
%
%
An $M$-augmenting path~\cite{papadimitriou1998combinatorial} is a path that
starts and ends with vertices that are not adjacent to edges that belong to
$M$, and all other vertices in this path are adjacent to such edges. Moreover,
every second edge in the path belongs to $M$. It is easy to see that by
removing all the edges of such a path from $M$, and by adding those in this
path that were not in $M$, one increases the size of $M$ by 1. This technique
is used to improve the size of maximal matchings.



\noindent
\macb{Greedy Approach}
A traditional greedy approach~\cite{cormen2009introduction} is
also used in streaming settings.

\noindent
\macb{Linear Programming (LP)}
Some problems can be reduced to a linear
program~\cite{papadimitriou1998combinatorial} which also provides a solution to
the initial problem.


\noindent
\macb{Local Ratio}
The local ratio technique~\cite{bar2004local} is an approach for solving
combinatorial optimization problems and is related to linear programming.

\begin{table}[hbtp]
%
\footnotesize
%
\begin{minipage}{\columnwidth}
\begin{center}
\def\arraystretch{1.4}
\begin{tabular}{llllll}
\toprule
\textbf{Reference} & \textbf{Model} & \textbf{Determinism} & \textbf{Technique} & \textbf{Matching type} &  \\ \midrule
\cite{feigenbaum2005graph} & Insertion-Only & Deterministic & - & Cardinality & \\
\cite[Theorem 6]{konrad2012maximum}  & Insertion-Only & Deterministic & Augmenting Paths & Cardinality &  \\
\cite[Theorem 2]{konrad2012maximum}  & Insertion-Only & Deterministic &  Augmenting Paths & Cardinality &  \\
\cite[Theorem 1.1]{kapralov2014approximating}  & Insertion-Only & Deterministic & Local Algorithm & Cardinality &  \\
\cite[Theorem 1]{feigenbaum2005graph}  & Insertion-Only & Deterministic &  Augmenting Paths & Cardinality &  \\
\cite[Theorem 19]{ahn2011linear}  & Insertion-Only & Deterministic & LP & Cardinality &  \\
\cite[Theorem 5]{konrad2012maximum} & Insertion-Only & Deterministic &  Augmenting Paths & Cardinality &  \\
\cite[Theorem 1]{konrad2012maximum} & Insertion-Only & Deterministic &  Augmenting Paths & Cardinality &  \\
\cite[Theorem 4]{konrad2012maximum} & Insertion-Only & Randomized &  Augmenting Paths & Cardinality &  \\
\cite{karp1990optimal} & Online & Randomized & - &  Cardinality &  \\
\cite[Theorem 20]{goel2012communication}  & Vertex-Arrival & Deterministic & - &  Cardinality &  \\
\cite[Theorem 2]{kapralov2013better} & Vertex-Arrival & Deterministic & - &  Cardinality &  \\
\cite{chitnis2016kernelization} & Dynamic Graph Stream & Randomized & General Sampling &Cardinality &  \\
\cite{chitnis2016kernelization} & Dynamic Graph Stream & Randomized & General Sampling &Cardinality &  \\
\cite[Theorem 1]{assadi2016maximum} & Dynamic Graph Stream & Randomized & $L_0$ Sampling &Cardinality &  \\
\midrule
\cite[Theorem 2]{feigenbaum2005graph} &  Insertion-Only & Deterministic & Greedy & Weighted &  \\
\cite[Theorem 3]{mcgregor2005finding} &  Insertion-Only & Deterministic &  Greedy &  Weighted &  \\
\cite[Theorem 3]{mcgregor2005finding} &  Insertion-Only & Deterministic &  Greedy &  Weighted &  \\
\cite{zelke2012weighted} &  Insertion-Only & Deterministic & - & Weighted &  \\
\cite{epstein2011improved} &  Insertion-Only & Deterministic & - & Weighted &  \\
\cite{grigorescu2016streaming} &  Insertion-Only & Deterministic & - &  Weighted &  \\
\cite{paz20172+} &  Insertion-Only & Deterministic & Local Ratio & Weighted &  \\
\cite{ghaffari2017space} &  Insertion-Only & Deterministic & Local Ratio & Weighted &  \\
\cite[Section 3.2]{feigenbaum2005graph} &  Insertion-Only & Deterministic & - & Weighted &  \\
\cite[Theorem 28]{ahn2011linear} &  Insertion-Only & Deterministic & LP & Weighted &  \\
\cite[Theorem 22]{ahn2011linear} &  Insertion-Only & Deterministic & LP & Weighted &  \\
\cite[Theorem 22]{ahn2011linear} &  Insertion-Only & Deterministic & LP & Weighted &  \\
\midrule
\textbf{\cite{crouch2014improved}} &  Insertion-Only & Deterministic & - & Weighted &  \\
  \bottomrule
\end{tabular}
\end{center}
\end{minipage}
\caption{(\cref{sec:other-algs}) \textbf{Comparison of algorithms for maximum matching}. 
%
%
\textbf{Model}: model used to construct a given algorithm,
\textbf{Determinism}: whether a given algorithm is deterministic or randomized,
\textbf{Technique}: used general technique (see~\cref{sec:general-techniques});
}
\label{tab:algs-2}
\end{table}

\subsubsection{Maximum Cardinality Matching}
We start with algorithms for MCM algorithms.
%
%
%
A simple \emph{maximal} matching can be obtained by iterating over edges in
some arbitrary order and greedily adding an edge only if both its endpoints are
not used yet; this scheme requires $O(n)$ space~\cite{feigenbaum2005graph}.
Since every maximal matching is a $1/2$-approximation to a maximum matching,
this scheme leads to a $1/2$-approximation for the maximum cardinality
matching. For a long time, this approach was used to derive the best
approximation of a maximum matching, using only one pass.

Konrad et al.~\cite{konrad2012maximum} present a variety of algorithms, taking
either one or two passes over the input stream of edges. The general idea is to
simulate a three pass algorithm. The original three pass algorithm relies on
the refinement of a maximum matching in a bipartite graph using $M$-augmenting
paths, as already used by Feigenbaum et al.~\cite[Theorem
1]{feigenbaum2005graph}.

Kapralov et al.~\cite{kapralov2014approximating} simulate a local distributed
algorithm in the semi-streaming model. The local algorithm is able to
distinguish graphs with an $\Omega(n)$ size matching from the graphs having
no $n /\polylogshort(n)$ size matching. This approach is transformed
into a one-pass semi-streaming algorithm, requiring only $O(\polylogshort(n))$
space.  $O(\log(n))$ many instances of this algorithm are executed in parallel,
%
%
resulting in an $O(\polylogshort(n))$-approximation.

For bipartite graphs, $(1-1/e)$-approximations are possible in both the online
\cite{karp1990optimal} and vertex model~\cite{goel2012communication}. Both
assume that both vertex classes belonging to sets $U, W$ of the bipartite graph
$G = (U, W, E)$ have the same size, and that one set of the vertices is known
beforehand (say $U$).  The other set (say $W$) is then streamed in and at the
same time the edges are revealed to the other set. However, the two approaches
differ by the fact that one is online, so must make a decision as soon as the
edges arrive, and the other can defer the decision to a later point in time.
Additionally, the algorithm of Goel et al.~\cite{goel2012communication} is
deterministic. A refinement of Kapralov's scheme~\cite{kapralov2013better}
allows multiple passes on the input and also achieves a $(1-1/e)$-approximation
for one pass. Differently, the $|U| = |W|$ constraint is not mentioned.

In the dynamic graph stream model, Chitnis et al.~\cite{chitnis2016kernelization}
present an exact approximation using $\tilde{O}(k^2)$ space (where $k$ is the
size of the matching $|S^*|$), requiring only one pass. The approach is refined
for $(1/\alpha)$-approximative matchings, using $\tilde{O}(n^2 /\alpha^3)$
  memory. Both algorithms rely on a sampling primitive, which runs in parallel
  and is also applicable in the MapReduce~\cite{dean2008mapreduce} setting.

A one-pass algorithm in the dynamic graph stream model is presented by
Assadi et al.~\cite[Theorem 1.1]{assadi2016maximum}. The algorithm uses a
bipartite graph and two approximations of the matching as the input. Note that
one can run the algorithm for $O(\log(n))$ many estimates of the matching, to
determine the correct approximation value. The algorithm relies on
$L_0$-sampling to process the input and succeeds with probability of at least
$0.15$. Despite the fact that the algorithm runs for bipartite graphs only, by
choosing a random bipartition of the vertices it is possible to run the
algorithm for arbitrary graphs, reducing the approximation by a factor of at
most two.

Works on maximum matchings in low arboricity graphs also
exist~\cite{cormode2016sparse, mcgregor2018simple, mcgregor2016planar}.

\subsubsection{Maximum Weighted Matching} \label{sec:mwm}

Feigenbaum et al.~\cite{feigenbaum2005graph} presented the first
$6$-approximation in 2005. In the same year, the bound was improved to
$5.82$~\cite{mcgregor2005finding}, which also allows a
$(2+\epsilon)$-approximation using a constant number of passes, assuming
$\epsilon$ is small. Note that both of these one-pass algorithms decide at edge
arrival, if the edge is kept or not by comparing the weight of the incoming edge to
some value that depends on the matching computed so far. Using so called
shadow-edges, which may be reinserted into the matching later
on~\cite{zelke2012weighted}, the approximation value can be reduced to $5.58$.
Epstein et al.~\cite{epstein2011improved} partition the input edges into
$O(\log(n))$ edge sets. For each set, a separate maximal cardinality matching
is computed. Finally, a greedy algorithm is applied to merge the
cardinality matchings. This randomized method allows a $(4.911 +
\epsilon)$-approximation, and can be derandomized by running all possible
outcomes of the randomized algorithm in parallel. Note that there is a constant
number of parallel executions for a fixed $\epsilon$. A similar approach is
used to lower the one-pass approximation to
$4+\epsilon$~\cite{crouch2014improved}: The algorithm reduces the maximum
weight matching problem to a polylog number of copies of the maximum
cardinality matching problem. At the end, a greedy merge step is applied to get
the final result. It is also proven that this
specific approach cannot provide a better approximation than $3.5+ \epsilon$.
This lower bound of $3.5+\epsilon$ was achieved two years
later~\cite{grigorescu2016streaming}. Recently, the $2+\epsilon$ approximation
ratio was achieved~\cite{paz20172+} using the local ratio
technique~\cite{bar1985local, bar2001unified}.
Ghaffari~\cite{ghaffari2017space} improved the algorithm and reduced the space
required from $O(n \log ^2(n))$ to $O(n \log(n))$. The proof is done
differently using a blaming-charging argument.

Different multi-pass algorithms exist: Feigenbaum et
al.~\cite{feigenbaum2005graph} noticed that multiple passes allow to emulate an
already existing algorithm~\cite{uehara2000parallel} solving the maximum
weighted matching problem. This approach uses only $O(n \log(n))$ space
resulting in an $(2 + \epsilon)$-approximation with $O(\log(n))$ passes. Ahn
et al.~\cite{ahn2011linear} rely on linear programming: given a graph $G =(V,
E, w)$, a suitable linear program is defined, which needs to be solved. The
Multiplicative Weights Update Meta-Method~\cite{arora2012multiplicative} is
used to solve the linear program. Different approaches are presented to lower
the amount of space and passes on the input stream.

\section{Maximum Matching on FPGA}
\label{sec:main}

We now describe the design and implementation of the substream-centric
MM for FPGAs.

\subsection{Overview of the Algorithm}
\label{sec:algorithm}

%

We start with a high-level overview of the MWM algorithm. A pseudo code is
shown in Listing~\ref{background:algorithm:mwm:sf}. For each edge, we iterate
in the descending order of $i$ over the $L$ substreams, identifying them by
their respective weights (Line~\ref{lst:mwm:l1}). The $i$-th substream weight
is given by $(1+\epsilon)^i$. For each maximum matching~$C_i$, we use a bit
matrix $MB$ to track if a vertex has an incident edge to \emph{ensure that $C_i$
remains a matching} (i.e., that no two vertices share an edge). Bits included in $MB$ are called \emph{matching bits}.
Bits in $MB$ associated with a vertex~$u$, \emph{the source vertex of a
processed edge}, ($u$-matching bits) determine if $u$ has an incident edge
included in some matching; they are included in column~$mb_u$ of matrix~$MB$.
Matching bits associated with vertex~$v$, \emph{the destination vertex of a
processed edge}, ($v$-matching bits) track the incident edges of $v$;
they are included in column~$mb_v$ of matrix~$MB$.
%
%
Since there are $L$ matchings and $n$ vertices, the bit matrix $MB$ is a matrix
of size $L \times n$. Furthermore, every matching stores its edges in a list.
If an edge is added, a flag is set to \texttt{true} to prevent that the edge is
added to multiple lists (Line~\ref{lst:mwm:l2}). This reduces the runtime of
the post-processing part, in which we iterate in the descending order over the
$L$ lists of edges to generate the $(4+\epsilon)$-approximation to the maximum
weighted matching.
%
%

\subsubsection{Time \& Space Complexity}
The space complexity is $O(n L)$ to track the matching bits, and $O(\min(m,
n/2) L \log(n))$ to store the edges of $L$ maximum matchings. The time
complexity is $O(m L)$ for substream processing on the FPGA and $O(n L)$ for
substream merging on the CPU, resulting in the total complexity of $O(m L + n L)$.

%
%


\begin{lstlisting}[float=t, caption={(\cref{sec:algorithm}) \textbf{The high-level overview of the substream-centric
Maximum Weighted Matching algorithm}, based on the scheme by Crouch and Stubbs~\cite{crouch2014improved}},label=background:algorithm:mwm:sf]
@\tikzmark{io}@@\tikzmarkin{col-0-0}(11,0.3)(-0.05,-0.2)\hspace{-0.5em}@//@\textbf{\underline{Input:}}@ $\epsilon$, $E$, $L$. @\textbf{\underline{Output:}}@ T (a $(4+\epsilon)$-approximation of MWM).   @\myboxX{\textbf{ I/O }}@@\tikzmarkend{col-0-0}@

@\tikzmarkin{col-0-1}(11,0.25)(-0.05,-5.2)@//@\textbf{\underline{PART 1 (Stream processing):}}@ compute $L$ maximum matchings
C: List of Lists; //$L$ lists to store edges in $L$ substreams
MB: Matrix; //The matching bits matrix of size $L \times n$
substream_weights: List; //The list of substream weights;
//substream_weights[i] = $(1+\epsilon)^i$.
has_added: bool; //Controlling adding an edge to only one MCM
foreach(WeightedEdge e : $E$) {
  has_added = false;
  for(i = $L-1$; i >= $0$; i--) { @\label{lst:mwm:l1}@
    if(e.weight >= substream_weights[i]) {
      if(!MB[e.u][i] && !MB[e.v][i]) {
        MB[e.u][i] = 1; MB[e.v][i] = 1; 
        if(!has_added) {//Add e only once to the matchings
          C[i].add(e); has_added = true; @\label{lst:mwm:l2}@
} } } } }                                                 @\myboxX{\textbf{ FPGA }}@@\tikzmarkend{col-0-1}@

@\tikzmarkin{col-0-2}(11,0.25)(-0.05,-3.35)@//@\textbf{\underline{PART 2 (Post processing):}}@ combine $L$ matchings into a MWM
T: List; //A list with the edges of the final MWM
tbits: List; //An array containing the matching bits of T
for(i = $L-1$; i >= $0$; i--) { 
  foreach(WeightedEdge e : C[i]) {
    if(!tbits[e.u] && !tbits[e.v]) {
      tbits[e.u] = 1; tbits[e.v] = 1;
      T.add(e);
} } }
return T;                                                  @\myboxX{\textbf{ CPU }}@@\tikzmarkend{col-0-2}@
\end{lstlisting}

\subsubsection{Reducing Data Transfer with Matching Bits Storage}
We assume that the input is streamed according to the CSR order corresponding
to the input adjacency matrix.
If we process a matrix row, we load the edges from DRAM
to the FPGA. Further, we can store the matching bits $mb_{u}$ of vertex $u$ in
BRAM on the FPGA, \emph{since they are reused multiple times {(temporal locality)}}. The matching bits of
$v$ are streamed in from DRAM. Since the matching bits for $v$ are not used
afterwards for the same matrix row, we write them back to DRAM. Using this
approach, we can process the whole graph row by row and need to store only the
$u$-matching bits in BRAM.
%
%


\subsection{Blocking Design for More Performance}
\label{sec:blocking-high-level}

\subsubsection{Problem of Data Dependency}
We cannot start processing the next row of the adjacency matrix until the last
matching bits of the previous row have been written to DRAM, because we might
require accessing the same $v$-matching bits again (read after write
dependency). 
%
%
In such a design, the waiting time required after each row could grow,
decreasing performance.  

\subsubsection{Solution with Blocking Rows}
We alleviate the data dependency by applying \emph{blocking}. 
%
%
We merge $K$ adjacent rows to become one stream; we call the merged stream of
$K$ rows an \emph{epoch}, and denote the $k$-th epoch (starting counting from
1) as $k$. There are $\ceil{n/K}$ epochs in total. To enable merging the rows,
we define a \emph{lexicographic ordering} over all edges. 


\subsubsection{Lexicographic Ordering}
Let a tuple $(u, v, w, k)$ denote an edge with vertices $u, v$, weight $w$,
and associated epoch $k = \floor{ (u-1) / K} + 1$. 
Then, the lexicographic ordering is given by: $(u_a, v_a, w_a, k_a) <
(u_b, v_b, w_b, k_b) $ iff $k_a < k_b  \lor ( k_a = k_b \land v_a < v_b ) \lor
( k_a = k_b \land v_a = v_b \land u_a < u_b)$; the edge weight is ignored. An
example is in Figure~\ref{fig:implementation:improved:input} (top). The
lexicographic ordering is implemented by a simple merging network.
%

\begin{lstlisting}[showstringspaces=false, float=t, caption={(\cref{sec:algorithm}--\cref{sec:alg-fpga-details})
The pseudocode of the substream-centric MWM algorithm, \textbf{enhanced with
the blocking optimization and a lexicographic ordering}.},
label=listing:implementation:improved:pseudocode]
@\tikzmarkin{col-1-0}(11,0.3)(-0.05,-0.175)@//@\textbf{\underline{Input and Output}}@: as in Listing @\ref{background:algorithm:mwm:sf}@.                         @\myboxX{\textbf{ I/O }}@@\tikzmarkend{col-1-0}@

@\tikzmarkin{col-1-1}(11,0.35)(-0.05,-8.4)@//@\textbf{\underline{PART 1 (Stream processing):}}@ compute $L$ maximum matchings
for(Epoch k = 1; k <= $\ceil{n/K}$; k++) {
  Load u-matching bits from DRAM into @double@-buffered BRAM
  Merge the $K$ rows of edges (loaded from DRAM into one stream S)
    with a merging network (Figure @\ref{fig:main_design}@), apply lexicographic order
  //Process each edge
  foreach(WeightedEdge e : S) {
    Matching bits requester loads matching bits (e.v) from DRAM@\label{listing:implementation:optimized:pseudocode:ptotal:l1}@
    //Apply the 8 stage pipeline for each edge
    @\textbf{Stage 1:}@ extract v-matching bits from a data chunk,
      determine BRAM address
    @\textbf{Stage 2:}@ load the matching bits @for@ e.u from BRAM
    @\textbf{Stage 3:}@ wait @for@ one cycle due to the latency of the BRAM
    @\textbf{Stage 4:}@ store the arriving BRAM data in a register, select
      the correct matching bits, compute el[i] = e.w = $(1+\epsilon)^i$
    @\textbf{Stage 5:}@ compute the matchings
    @\textbf{Stage 6:}@ write u-matching bits to BRAM, write
      v-matching bits to double-buffered BRAM if required
    @\textbf{Stage 7:}@ determine the least significant bit in te,
      store them in variable i
    @\textbf{Stage 8:}@ write the edge to DRAM at C[i]
      (if part of a matching), write v-matching bits to DRAM@\label{listing:implementation:optimized:pseudocode:ptotal:l2}@
  }
  Wait till all writes to DRAM are committed}             @\myboxX{\textbf{ FPGA }}@@\tikzmarkend{col-1-1}@

@\tikzmarkin{col-1-2}(11,0.3)(-0.05,-0.145)@//@\textbf{\underline{PART 2 (Post processing):}}@ As in Listing @\ref{background:algorithm:mwm:sf}@.                  @\myboxX{\textbf{ CPU }}@@\tikzmarkend{col-1-2}@
\end{lstlisting}

\subsubsection{Advantages of Blocking}
%
%
At the end of each epoch, $v$-matching bits are written to DRAM. This reduces
the number of such transfers from $n$ to $n/K$. Moreover, if edges in different
rows share the same $v$-matching bits, only one load from DRAM is required.
Finally, $u$-matching bits can be kept in BRAM, since they are reused multiple
times. 


\subsubsection{Further Optimizations}
To achieve a performance of \emph{up to one processed edge per cycle}, we
pipeline the processed edges, we distribute the $u$-matching bits over multiple
BRAMs (to facilitate reading data from different addresses), and we double
buffer $u$-matching bits to reduce latencies.

\begin{figure}[b!]
  \centering
  \includegraphics[width=0.8\textwidth]{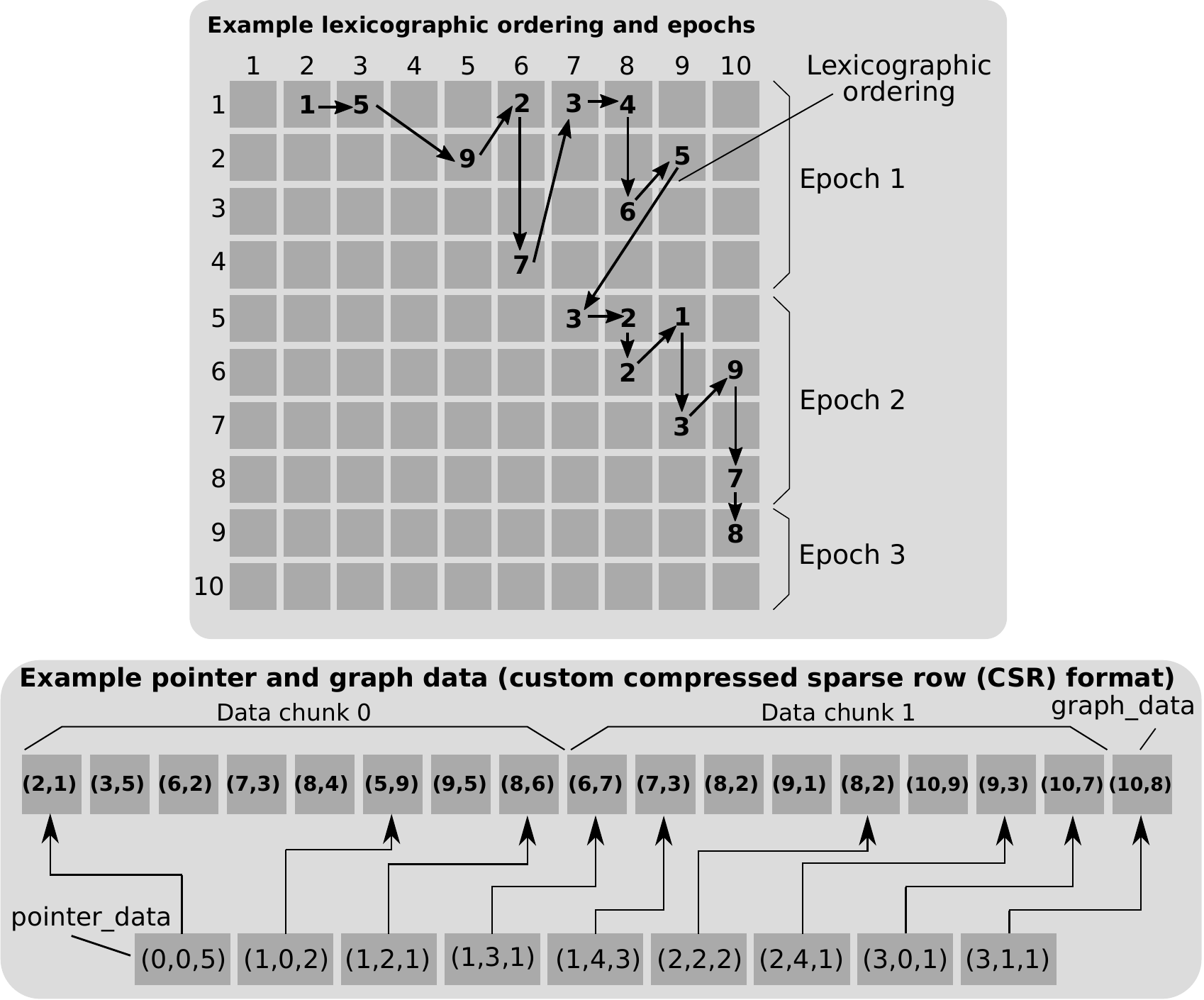}
%
\caption{An example input adjacency matrix,
its annotated lexicographic ordering
illustrated by arrows ($K=4$), and
and its custom compressed sparse row
(CSR) format. The entries of the adjacency matrix denote the weight of an
edge.}
%
  \label{fig:implementation:improved:input}
\end{figure}

\begin{figure}[hbtp]
\vspace{-1em}
  \centering
  \begin{turn}{90}
  \includegraphics[width=1.4\textwidth]{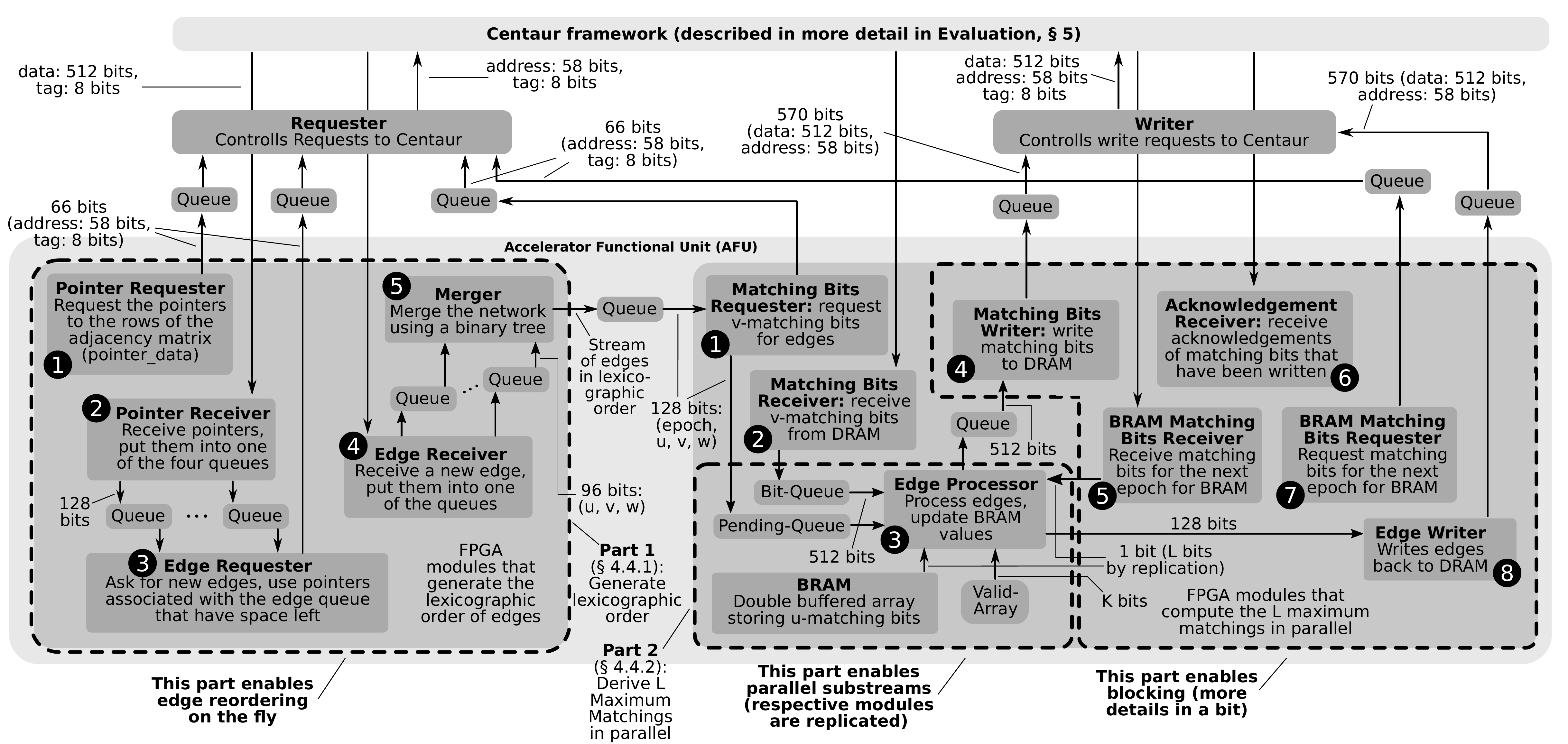}
\end{turn}
\vspace{-1em}
\caption{
  (\cref{sec:alg-fpga-details}) The \textbf{interaction of the FPGA modules} to approximate
MWM. 
For clarity, the State Controller is omitted.
The wires of incoming data from Centaur consists of 512 bits for data and 8 bits for tag.
All modules are connected using AXI interfaces. {All valid bits are omitted.
The merger network is in Figure~}\ref{fig:merger}.
%
%
}
%
  \label{fig:main_design}
\end{figure}

\begin{figure}
  \centering
  \includegraphics[width=0.6\textwidth]{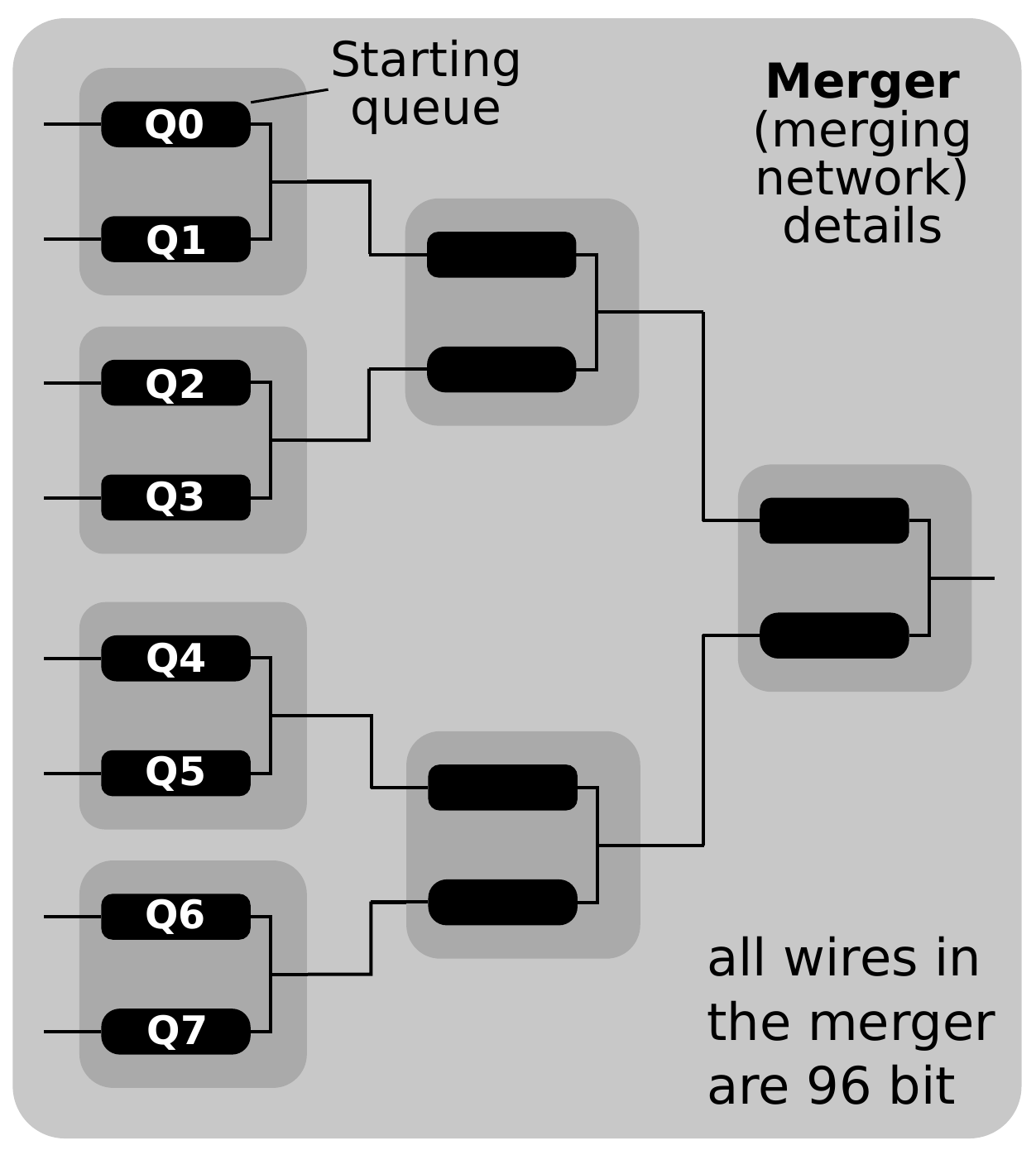}
%
\caption{
  (\cref{sec:alg-fpga-details}){ The \textbf{merger network} 
from Figure~}\ref{fig:main_design} for $K=8$.
%
%
}
%
  \label{fig:merger}
\end{figure}

\subsection{Input and Output Format}

The input to the FPGA algorithm is a custom variant of the Compressed Sparse
Row (CSR) format. An example is given in
Figure~\ref{fig:implementation:improved:input} (bottom).
The format has two parts: The \texttt{pointer\_data} and the
\texttt{graph\_data}.
First, the \texttt{pointer\_data} stores information about the start and end of
each row of the adjacency matrix. An entry contains three parts: the ID of the
\emph{data chunk} with information about where the first edge is stored, the
data chunk offset denoting the offset of the first edge from the start of the
data chunk, and the number of associated edges (a data chunk refers to data of
a given size at an aligned memory address). Each entry uses 32 bits, making an
entry of the \texttt{pointer\_data} 96 bits. We fit five entries (480 bits) in
a data chunk.
%
%
Second, the \texttt{graph\_data} is a stream of edges. One entry consists of
the column index and the edge weight.  The row identifier is given by the
corresponding entry in the \texttt{pointer\_data}. One \texttt{graph\_data}
entry requires 64 bits, allowing to store eight edges in a data chunk.

Our custom data layout has
different advantages over the usual CSR format. First, a single entry of the
\texttt{pointer\_data} already gives all required information about the start
and length of the row of the adjacency matrix. This entails some redundancy
compared to the
traditional CSR, but only requires one load from DRAM to resolve a given edge.
%
%
Further, CSR splits the
column indices and values. We merge them together in one stream,
reducing the number of random accesses.
%
%

The output of the FPGA consists of $L$ substreams of edges. The $i$-th
stream contains edges of the maximum matching $C_i$. 
We use 128 bits for each edge: 32 bits each for the vertex
IDs, the edge weight, and the assigned index $i$ of the maximum matching (which
could be omitted).  A single data chunk therefore contains four output edges.

\subsection{Details of Processing Substreams on FPGA}
\label{sec:alg-fpga-details}

We explain the interaction of the FPGA modules 
dedicated to generating the lexicographic ordering (Part~1) 
and computing the maximum matchings (Part~2); see
Figure~\ref{fig:main_design}, {Figure~}\ref{fig:merger}, and
Listing~\ref{listing:implementation:improved:pseudocode}.



\subsubsection{Generating Lexicographic Ordering}
As input to the FPGA, we get the address pointing to the start of
\texttt{pointer\_data}, the number of vertices $n$, the number of edges $m$, a
pointer $p_{out}$ where we write the output to, and an offset value $o$ to
distinguish the $L$ output streams (start of output stream $i$ is at $p_{out} +
i \cdot o$). The \textbf{pointer requester} is responsible for requesting the
data chunks holding the \texttt{pointer\_data}. The requested pointers arrive
at the \textbf{pointer receiver}. Given a data chunk, the pointer receiver unwraps the five 
pointers, and passes them to the \textbf{edge requester}.
%
%
The
\texttt{pointer\_data} from the pointer receiver is passed to four different
queues $Q_0, Q_1, Q_2, Q_3$, where every queue gets a subset of the pointers
dependent on $K$. Assume for simplicity that the vertex IDs start at 0 and $(K
\mod 4) = 0$. Then, to be precise, given a pointer $p(u)$ pointing to row $u$,
we assign $p(u)$ to $Q_i$ if $ (u \mod K) \geq K/4 \cdot i \land (u \mod K ) <
K/4 \cdot (i+1)$.  For example, with $K=16$, $Q_0$ stores $p(u)$ with $u=0, 1,
2, 3, 16, 17, 18, 19, 31 \ldots$, and $Q_1$ stores $p(u)$ with $u = 4, 5, 6, 7,
20, \ldots$. The \texttt{pointer\_data} is loaded from the queues into a
BRAM array $BP$ of size $K$, where every entry holds two pointers ($2K$ pointers
are therefore stored in total). If an entry $i$ of $BP$ has pointers $p(u')$
and $p(u'')$, it holds that $i = (u' \mod K) = (u'' \mod K)$ and $p(u'')$
requests edges for an epoch after $p(u')$.  Therefore, only the first pointer
in an entry is valid to use and we have random access to $K$ valid pointers in
total. To describe the mechanism that determines the selection of the next
pointer to request new edges, we first inspect further processing steps.


The \textbf{edge receiver} gets data chunks containing \texttt{graph\_data}
from the framework and unwraps them (we use the Centaur framework~\cite{owaida2017centaur}
to access main memory independently of the CPU).
Information regarding the offset and
number of edges which are valid for a data chunk request is also passed from
the edge requester to the edge receiver.
%
%
Next, an edge $e = (u, v, w)$ is passed from the edge receiver to the
\textbf{merger}.  There, the edge is inserted in a \emph{starting queue} (with ID~$(u \mod K)$).
%
%
The merger
merges the $K$ streams in lexicographic ordering.  It
consists of a series of merging elements, where each element has two input
queues and an output port.  The element compares edges in its queues and
outputs the edges according to the lexicographic ordering.  The merging
elements form a binary tree, such that for a given $K$,
there are $K/2$ starting elements with $K$ starting queues in total.

The edge requester can observe the size of the starting queues of the merger.
It operates in two modes to determine a pointer to new edges. In mode~1,
one selects a pointer $p(u)$ from queue~$Q_i$ as the next candidate if the
corresponding starting (merger) queue $(u \mod K)$ does not overflow, and
store the pointer in BRAM~$BP$ at position $(u \mod K)$. If mode~1 
fails (for example, if there is no empty space at the appropriate position in BP), then mode~2 selects the pointer according to the merger
starting queue which has the least amount of edges. Note that the edge
requester also takes the requests which are in flight into account to predict
the future size of the starting queue. This approach ensures that the
merger queues do not overflow and their load is balanced.

For a row~$u$ which has no edges, a special
information is passed from the edge requester to the edge receiver. It then
inserts an artificial edge in the merger. This allows to overcome
problems, where a merging element waits for new input, but does not
receive any, since the adjacency matrix row is empty. The merging
network filters these edges at the output port (they are not passed
on).



\subsubsection{Deriving $L$ Maximum Matchings}
The stream in lexicographic ordering is passed to the \textbf{matching bits
requester}.  This module requests the $v$-matching bits from DRAM.  It can only
operate when the bits of the epoch before have been acknowledged.  Also, it
only processes edges belonging to the current epoch which is defined by the
\textbf{state controller}.  The requested data is received in the
\textbf{matching bits receiver}.  It passes the full data chunk to the
\textbf{edge processor}.  Using the matching bits and the ordered stream of
edges, the edge processor computes the $L$ maximum matchings in parallel in an
8-stage pipeline (Listing~\ref{listing:implementation:improved:pseudocode},
Lines
\ref{listing:implementation:optimized:pseudocode:ptotal:l1}--\ref{listing:implementation:optimized:pseudocode:ptotal:l2}).
In Stage~1, $v$-matching bits for a given edge are extracted from a data chunk.
Further, the address of the $u$-matching bits in BRAM is computed.
Since the more up-to-date $v$-matching bits might also be stored in BRAM, this address is also
determined. In Stage~2, read requests to fetch the matching bits from BRAM are
issued. Stage~3 only waits one clock cycle for BRAM to return the data. In
Stage~4, the BRAM data arrives and is stored in a register. The stage also
decides if $v$-matching bits are taken from the data chunk or from BRAM.
Further, the stage computes the matching value $te$ indicating if an edge
$e=(u, v,w)$ belongs to substream $E_i$; $te[i] = w \geq (1+\epsilon)^i$ for $i
\in  \{0, \ldots, L-1\}$. In Stage~5, the actual matching is computed. As 
the BRAM data from Stage~4 may already be obsolete,
%
%
the computed values are also stored in registers for instant access in the next
cycle. The result is passed to Stage~6, in which the updated $u$-matching bits
(and if required also the $v$-matching bits) are written back to BRAM. In
Stage~7, the maximum matching with the highest index, to which the edge is
assigned, is determined. 
%
%
Finally, Stage 8 passes the edge to the \textbf{edge writer} to write it back
to DRAM (if the edge is used in a matching) and also passes the updated
$v$-matching bits to the \textbf{matching bits writer} for writing back to
DRAM.

The BRAM storing the $u$-matching bits is double buffered. While the first
BRAM buffer is used in the edge processor, the matching bits for the next epoch
are loaded from DRAM to the second BRAM buffer. Since an epoch can alter the
$u$-matching bits required for the next epoch, we write the according updates
also in the double buffered BRAM if required. To prevent that stale data from
DRAM overwrites the more up-to-date data, we use a register (the valid-array)
as flag. After an epoch, the access is redirected to the BRAM containing the
loaded data. The \textbf{BRAM matching bits requester} requests the according
data from DRAM, and the \textbf{BRAM matching bits receiver} unwraps the data
chunks. It passes the data to the edge processor. There, Stage 6 checks for
data from the BRAM matching bits receiver and updates the according entry in
the BRAM.

The \textbf{acknowledgement receiver} tracks the count of write
acknowledgements from the framework and determines if all $v$-matching bits are
committed to DRAM when an epoch ends.  When all edges from the epoch are
processed, the state controller indicates the start of the next epoch.



\subsection{Substream Merging on the CPU}


After the $L$ MCMs are written to DRAM, the CPU inspects them in the decreasing
order to compute the final maximum matching $(4+\epsilon)$-approximation.
{This part is a simple greedy scheme that exposes little parallelism, thus
we execute it on the CPU. It takes $O(L n)$ time and $O(L n)$ work.}

\subsection{{Summary of Optimizations}}

{In Figure~}\ref{fig:main_design}{, we also use dashed rectangles to
illustrate which modules are responsible for the most important optimizations:
edge reordering on the fly, parallel substreams (pipelining), and blocking.
Modules responsible for pipelining are appropriately replicated.
}


\subsection{{Interactions with DRAM}}

We use the Centaur
framework~\cite{owaida2017centaur} as the interface to the
Accelerator Functional Unit (AFU), the custom FPGA implementation, allowing to
access main memory independently of the CPU. 
%
%
Centaur consists of a software and a
hardware part. The software part allows to start and stop hardware functions,
to allocate and deallocate the shared memory, and pass input parameters to the
FPGA. The hardware part is responsible for bootstrapping the FPGA, setting up
the QPI endpoint, and handling reads and writes to the main memory.
{In our design, we use dedicated arbiter modules for all read and write
requests to Centaur: the requester and the writer. The requester has four
queues. The pointer requester, the edge requester, the matching bits requester, and the BRAM
matching bits requester can all write the DRAM address (from which a data chunk of
512 bits should be read) to these four requester queues. The requester uses a fixed priority order to send the requests to the
Centaur framework. Centaur provides an 8 bit tag to identify the requests. For simplicity, fixed
tags are used for each of the four modules emitting requests. The modules listening for incoming
data (pointer receiver, edge receiver, matching bits receiver, and BRAM matching bits receiver)
process the data chunk of size 512 bits only when the tag matches the expected value.
The design relies on the FIFO behavior of Centaur, such that
requests with the same tag are not reordered.
The writer orchestrates the modules (the matching bits writer
and the edge writer) which issue writes to DRAM.
Similarly to the requester, the writer has a queue for each writing module (two in total) and it
uses a fixed priority order. The
modules issuing the requests use fixed tags. Note that data is written in chunks of 512 bits.
The acknowledgment receiver monitors
the Centaur interface for writes that have been written. This information is passed to the
state controller (not shown in Figure~}\ref{fig:main_design}{) to orchestrate the modules.
Since Centaur allows only to access data in chunks of 512 bits, the addresses passed to the
framework have 58 bits.}


\section{Evaluation}
\label{sec:evaluation}

We now illustrate the advantages of our hybrid (CPU+FPGA) MWM design and inspect
resource and energy consumption.  
\emph{For every benchmark, each tested algorithm was synthesized, routed, and
executed on the hybrid FPGA platform specified below}.

\subsection{Setup, Methodology, Baselines}

\subsubsection{Compared Algorithms}
Since to our best knowledge \emph{no MWM algorithms for FPGAs are available},
we compare our design to three state-of-the-art CPU implementations.
In total, we evaluate three CPU and two CPU+FPGA algorithms; see
Table~\ref{table:evaluation:algorithms}. First \ding{182}, we implement a
sequential CPU-only version of the substream-centric MWM, based on the scheme
by Crouch and Stubbs~\cite{crouch2014improved}, as presented in
Listing~\ref{background:algorithm:mwm:sf} (CS-SEQ). Second \ding{183}, we
parallelize the algorithm with OpenMP's \texttt{parallel-for} statement to
compute different maximum matchings in parallel (CS-PAR). Third \ding{184}, we
implement the algorithm by Ghaffari~\cite{ghaffari2017space} (G-SEQ) that
provides a $(2+\epsilon)$-approximation to MWM with time complexity of $O(m)$
and space complexity of $O(n \log(n))$ bits. Thus, this algorithm is optimal in
the asymptotic time and space complexity. We compare these three algorithms to
our optimized FPGA+CPU implementation, SC-OPT \ding{185}. Finally, we also tested SC-SIMPLE \ding{186},
a variant of our implementation that uses no blocking.
SC-SIMPLE delivers more performance than the comparison baselines but it is consistently outperformed by
SC-OPT, we thus usually exclude it for clarity of presentation.
{However, we use it in power consumption experiments to illustrate
how much additional power is used by the design optimizations in SC-OPT}.
%
%
%
\emph{To our best knowledge, \textbf{we report the first performance
data for deriving maximum matchings on the FPGA}}.


\subsubsection{Implementation Details and Reproducibility}
We implement our algorithms {in Verilog} on a hybrid CPU+FPGA system using the Centaur
framework~\cite{owaida2017centaur}.
The modules outlined in Figure \ref{fig:main_design} are connected using AXI interfaces. 
{To facilitate reproducibility and interpretability~}\cite{hoefler2015scientific}{, we
make the whole code publicly available\footnote{https://spcl.inf.ethz.ch/Parallel\_Programming/Matchings-FPGA}.}


%

\begin{table}[h!]%

\begin{minipage}{\columnwidth}
\begin{center}
\begin{tabular}{lll}
  \toprule
  \textbf{Algorithm} & \textbf{Platform}  & \textbf{Time complexity}  \\ \midrule  
  Crouch et al. \cite{crouch2014improved} Sequential (CS-SEQ) & CPU &  $O(m L + n L )$  \\  
  Crouch et al. \cite{crouch2014improved} Parallel (CS-PAR) & CPU & $O(m L / T + n L)$  \\  
  Ghaffari \cite{ghaffari2017space} Sequential (G-SEQ) & CPU  & $O(m)$  \\  
   Substream-Centric, no blocking (SC-SIMPLE) & Hybrid & $O(m + n L^2)$ \\  
   Substream-Centric, with blocking (SC-OPT) & Hybrid & $O(m + n/K + n L)$   \\ 
 \bottomrule
\end{tabular}
\end{center}
\end{minipage}

\caption{(\cref{sec:evaluation}) Overview of \textbf{the evaluated MWM algorithm implementations}.}
%
\label{table:evaluation:algorithms}

\end{table}

%

\subsubsection{Setup}
We use Intel HARP~2~\cite{oliver2011reconfigurable}, a hybrid CPU+FPGA system.
It is a dual socket platform where one socket is occupied by an Intel Broadwell
Xeon E5-2680 v4 CPU~\cite{intelxeon} with 14 cores (28 threads) with up to 3.3
GHz clock frequency. Each core has 32 KByte L1 cache and there is 35 MByte L3
cache in total. An Arria-10 FPGA {(10AX115U3F45E2SGE3)} is in the other socket.
The used FPGA has speed grade~2~\cite{alterarria10}. It provides 55 Mbit in
2,713 BRAM units and 427,200 ALMs.  The FPGA is connected to the CPU by one QPI
and two PCIe links. The system runs Ubuntu 16.04.3 LTS with kernel 4.4.0-96 as
the operating system. All host code is compiled with gcc 5.4.0 and the -O3
compile flag.


\subsubsection{Datasets}
The input graphs are shown in Table~\ref{table:evaluation:graphs}.  We use both
synthetic (Kronecker) power-law graphs of size up to $n = 2^{21}, m = 48n$ from the
10th DIMACS challenge~\cite{dimacs10} 
%
%
and real world KONECT~\cite{konect} and SNAP~\cite{snapnets} graphs.
For unweighted graphs, we assigned weights uniformly at random with a fixed
seed. The value range is given by $[1, (1+\epsilon)^{L-1}+1]$.

\begin{table}[h!]%
%
\begin{minipage}{\columnwidth}
\begin{center}
\begin{tabular}{lllll}
 \toprule
 \textbf{Graph} & \textbf{Type} & \textbf{Reference} & $m$ & $n$ \\ \midrule
 Kronecker & Synthetic power-law & DIMACS 10 \cite{dimacs10} & $\approx$$48n$ & $2^k$; $k =16, \ldots, 21$ \\ 
 Gowalla & Social network & KONECT \cite{konect} & 950,327 & 196,591 \\ 
 Flickr & Social network & KONECT \cite{konect} & 33,140,017 & 2,302,925 \\ 
 LiveJournal1 & Social network & \specialcell[l]{SNAP \cite{snapnets}} & 68,993,773 & 4,847,571 \\ 
 Orkut & Social network & KONECT \cite{konect} & 117,184,899  & 3,072,441 \\ 
 Stanford & Hyperlink graph & KONECT \cite{konect} & 2,312,497 & 281,903 \\ 
 Berkeley & Hyperlink graph & KONECT \cite{konect} & 7,600,595 & 685,230 \\ 
 arXiv hep-th & Citation graph  & KONECT \cite{konect} &  352,807 &  27,770  \\ 
 \bottomrule
\end{tabular}
\end{center}
\end{minipage}

\caption{Selected used graph datasets. 
K$x$ denotes a
Kronecker graph with
$2^x$ vertices.}

\label{table:evaluation:graphs}
\end{table}

%

\sloppy
\subsubsection{Measurements}
The runtime is measured by \texttt{clock\_gettime} with parameter
\texttt{CLOCK\_MONOTONIC\_RAW}, allowing the nanosecond resolution. The runtime
of the FPGA implementations is determined by the Centaur framework. We execute
each benchmark ten times to gather statistics and we use box plot entries to
visualize data distributions.

\subsection{Scaling Size of Synthetic Graphs}
\label{sec:eval-weak-size-kron}

\begin{figure}[h!]

	\centering
		\includegraphics[width=0.49\textwidth]{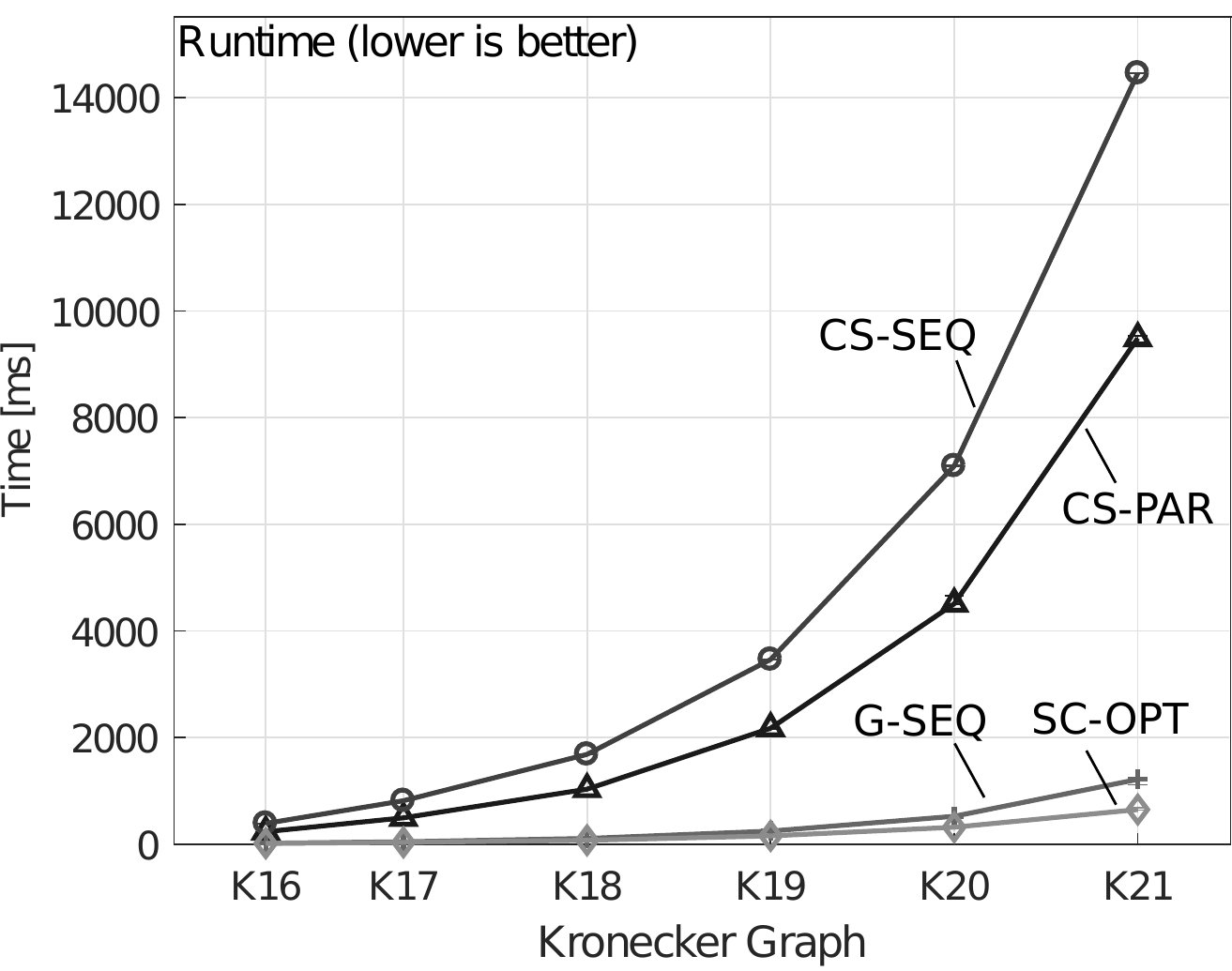}
		\includegraphics[width=0.47\textwidth]{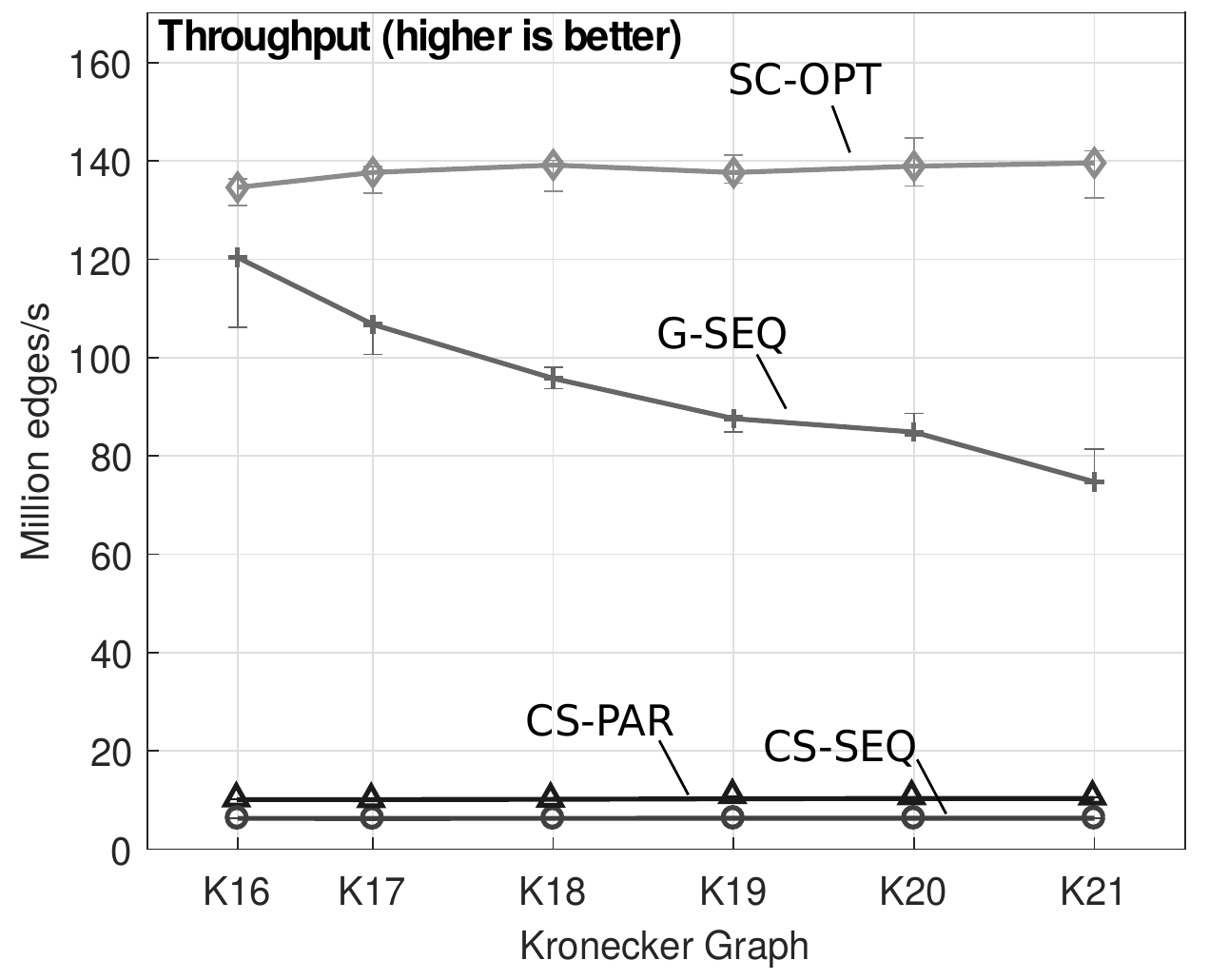}
%
\caption{(\cref{sec:eval-weak-size-kron}) \textbf{Influence of graph size $n$ on
performance} (synthetic power-law graphs).  $K=32, L=64, T=4, \epsilon = 0.1$.}
\label{fig:g-size-rmat}
\end{figure}

We first evaluate the impact from varying graph sizes (synthetic power-law
Kronecker graphs), for the fixed amount of parallelism (the \emph{weak scaling}
experiment).
The results are illustrated in Figure~\ref{fig:g-size-rmat}.
The throughput for
CS-SEQ and CS-PAR stays approximately constant below $\approx$12M edges/s. G-SEQ decreases in
performance as the graph size increases. We conjecture that this is due to the
increasing size of the hash map used to track pointers. This increases the
time for inserts and deletes, and might also require re-allocations to increase
the space. The performance for SC-OPT increases from $\approx$135M 
to $\approx$140M edges/s. This is because the initial (constant)
overhead (due to reading from DRAM) becomes less significant with larger graphs. 
%
%
%
\textbf{We conclude that the substream-centric SC-OPT beats 
comparison targets {for all considered sizes} of power-law Kronecker graphs.}


\subsection{Processing Different Real-World Graphs}
\label{sec:eval-weak-size-rw}

\begin{figure}[h!]
%
%
%
	\centering
		\includegraphics[width=0.495\textwidth]{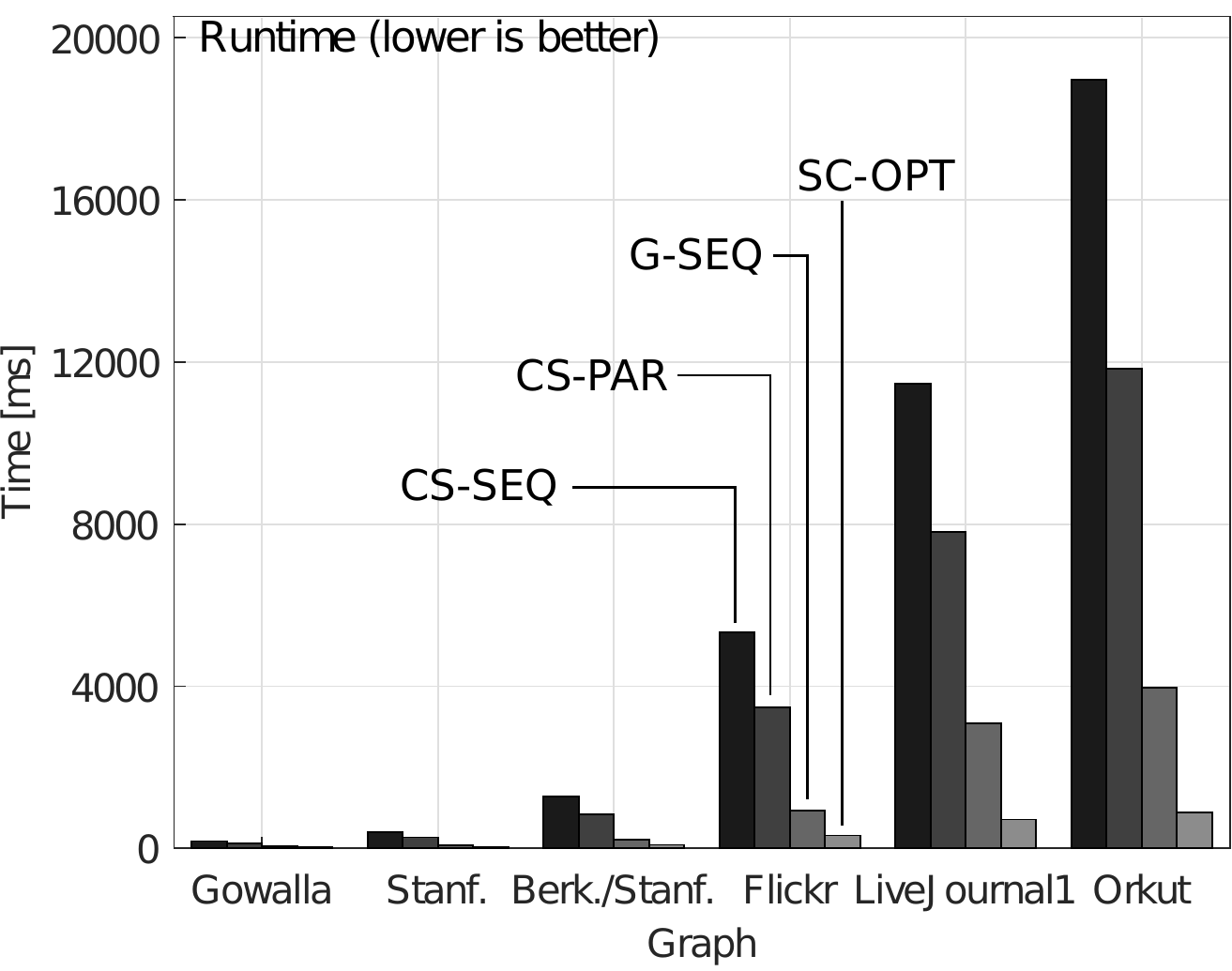}
		\includegraphics[width=0.48\textwidth]{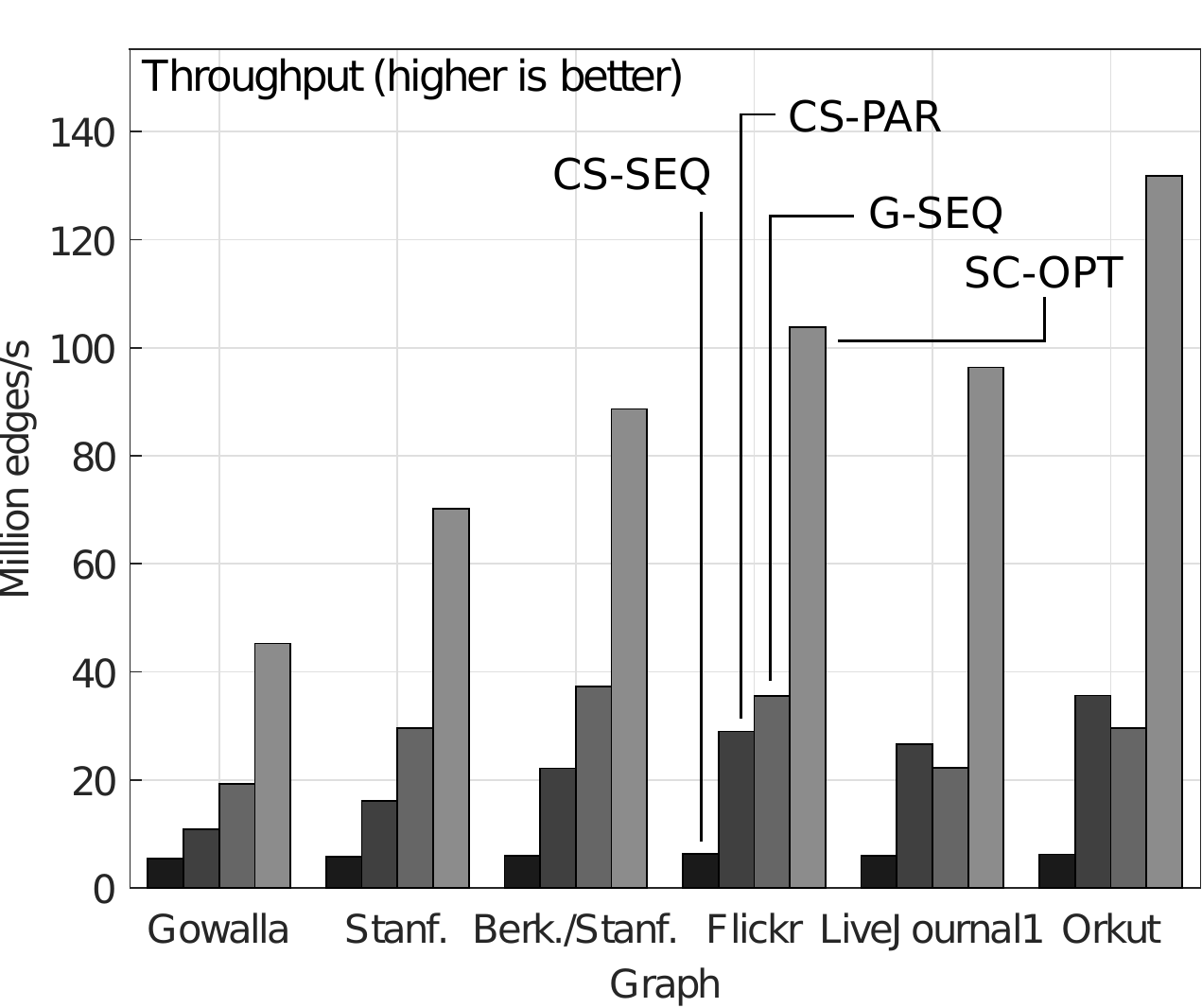}
%
\caption{(\cref{sec:eval-weak-size-rw}) \textbf{Influence of graph dataset $G$ on
performance} (real-world graphs). $K=32, L=64, T=4, \epsilon = 0.1$.}
\label{fig:g-size-rw}
\end{figure}

We next analyze the performance of the considered designs for different
real-world graphs; the results are illustrated in Figure~\ref{fig:g-size-rw}.
CS-SEQ and CS-PAR achieve sustained $\approx$3M edges/s and $\approx$10M
edges/s, respectively. The performance of SC-OPT is $\approx$45M edges/s for
small graphs due to the initial overhead of reading data from DRAM.  Compared
to the experiment with Kronecker graphs, the performance of both SC-OPT and
G-SEQ is lower for all graphs except Orkut. The reason is the average vertex
degree: it equals $\approx$48 in Kronecker graphs compared to $\approx$14 in
Flickr and LiveJournal1. If the ratio is high, G-SEQ can drop many edges
without further processing in an early phase. This reduces expensive updates to
the hash map and lists. For SC-OPT, the
waiting time (of data dependencies) 
%
%
lowers the performance. 
Still, \textbf{substream-centric SC-OPT ensures highest performance for all
considered real-world graphs}.
%

\subsection{Scaling Number of Threads $T$}
\label{sec:strong-scaling}

\begin{figure}[h!]
%
	\centering
		\includegraphics[width=0.495\textwidth]{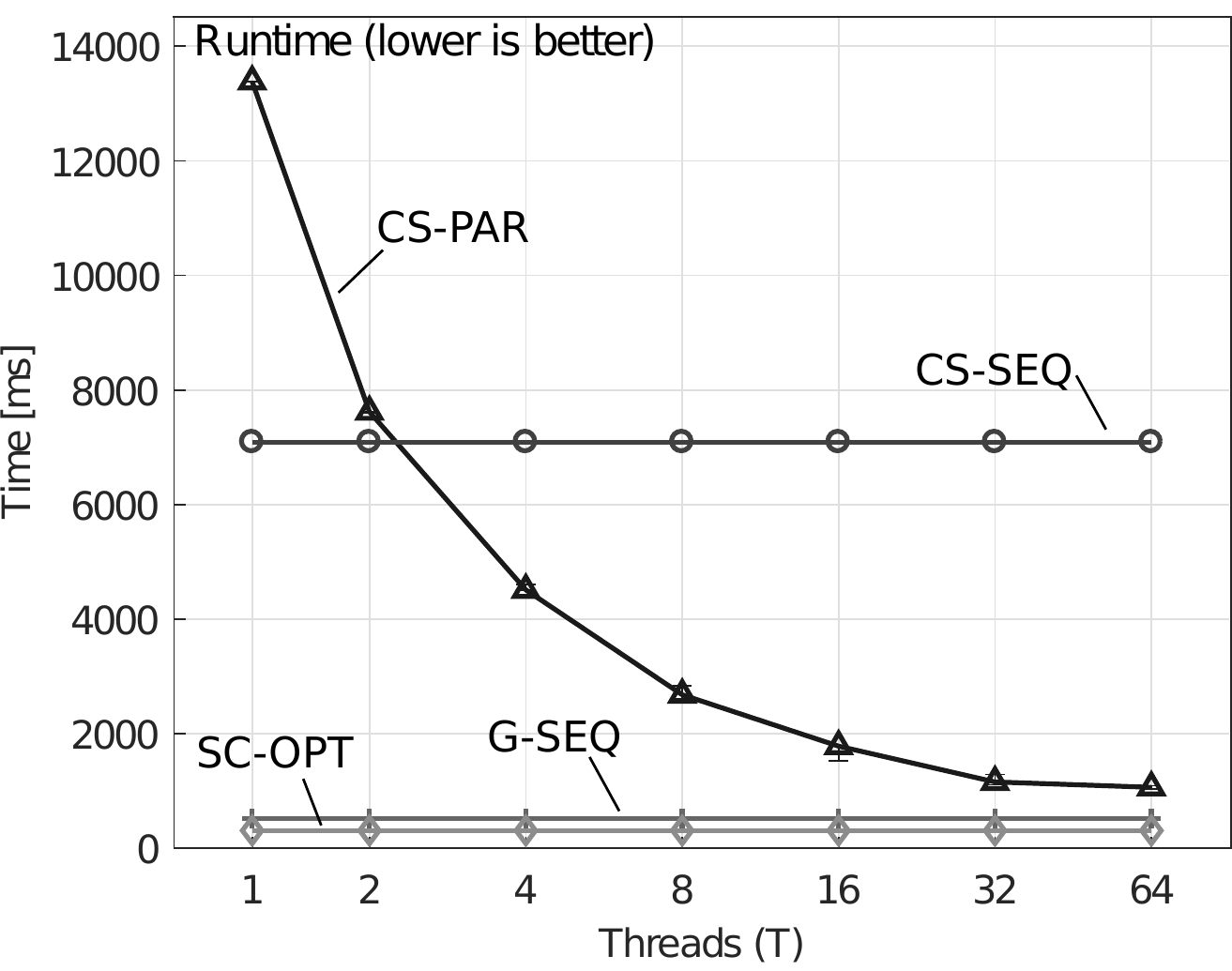}
		\includegraphics[width=0.48\textwidth]{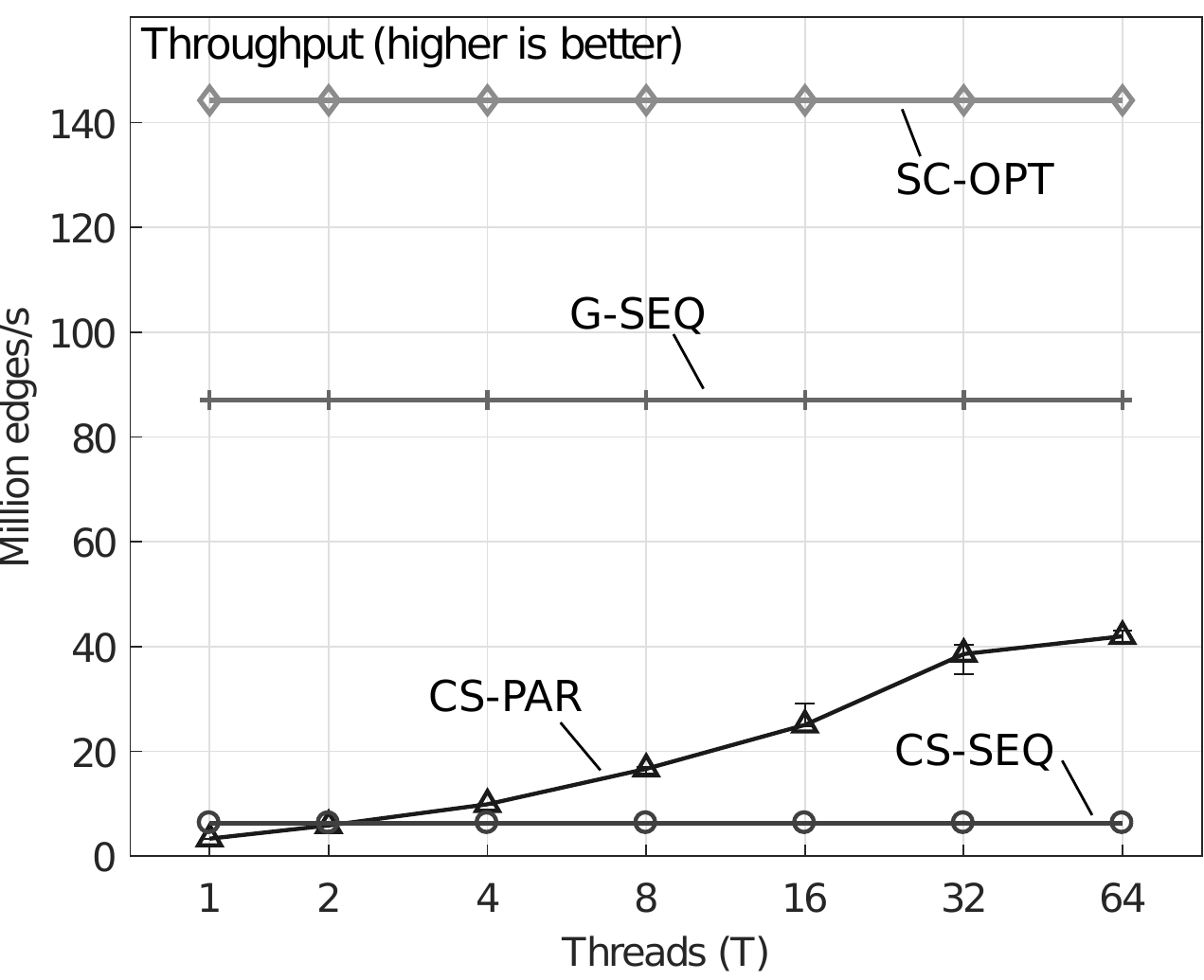}
%
	\caption{(\cref{sec:strong-scaling}) \textbf{Influence of the number of threads $T$ on performance}.
  The input graph is Kronecker with $n = 2^{20}, K = 32, L = 64, \epsilon = 0.1$.}
  \label{fig:strong-scaling}
\end{figure}

In the CPU versions, one can compute in parallel different maximum matchings in
SC-PAR using $T$ threads. In the following, we run a \emph{strong scaling}
experiment (fixed graph size, varying $T$) for a power-law Kronecker graph.
Figure~\ref{fig:strong-scaling} illustrates the results. 
Since G-SEQ and CS-SEQ are not multi-threaded, they do not scale with $T$.  The
parallelized CS-PAR reaches up to $\approx$40M edges/s, a $\approx$6$\times$
improvement over the sequential version, and an $\approx$14$\times$ improvement
over the parallel version with one thread. Therefore, the algorithm is still
$\approx$3$\times$ slower than SC-OPT which achieves up to $\approx$140M
edges/s on the K20 Kronecker graph.  Scaling is
limited since the parallel version takes $L$ passes over the stream, whereas
the other CPU algorithms process the input in one pass. The bandwidth usage of
the parallel version with $T = 64$ threads is $\approx$32 GB/s ($\approx$44M
edges and $64$ passes in one second), assuming no data sharing. Note that we only
parallelize the stream-processing part which computes the $L=64$ maximum
matchings.  However, as our analysis shows that the post-processing part takes
$<$1\% of the computation time of the maximum matching,
parallelization of post-processing would provide hardly any benefit.
We conjecture that the scaling of SC-PAR stops due to bandwidth limitations and
the limited computational resources of 14 cores.
%
%
Finally, \textbf{SC-OPT is the fastest regardless of $T$ used by other schemes}.

\subsection{Approximation Analysis}
\label{sec:acc}

\begin{figure}[h!]
	\centering
		\includegraphics[width=0.49\textwidth]{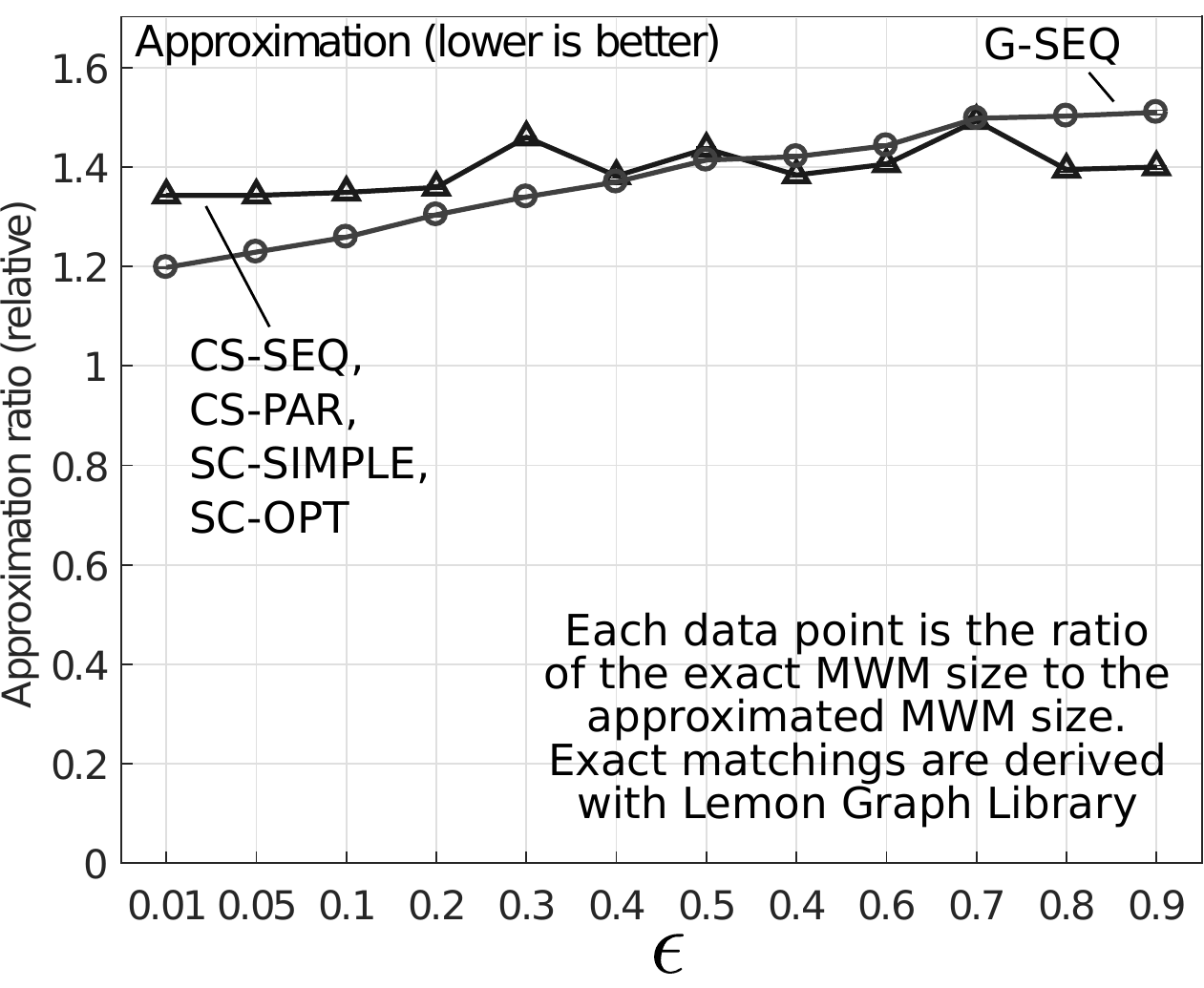}
		\includegraphics[width=0.49\textwidth]{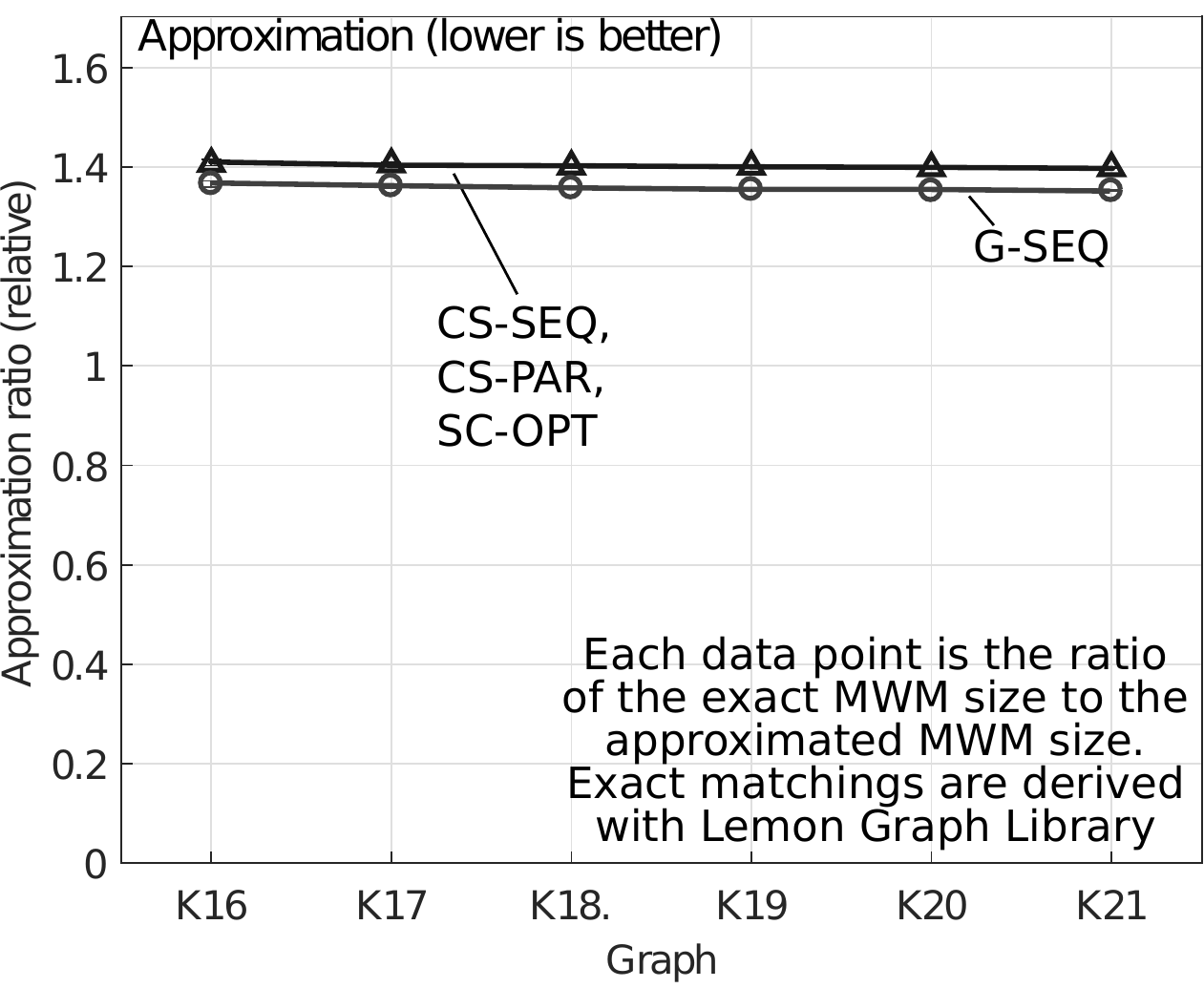}
%
	\caption{(\cref{sec:acc}) \textbf{Approximation analysis}.
The input graph is Kronecker with $n = 2^{19}$ (left).
$L=128$ and $\epsilon = 0.1$ (right).
{Explanation of the approximation ratio is provided in Section~2.}
} 
\label{fig:approx}
		
\end{figure}

We briefly analyze how well in practice SC-OPT approximates the exact MWM. The
results are in Figure~\ref{fig:approx} (SC-OPT, SC-SIMPLE, CS-SEQ, and CS-PAR
produce the same results). The accuracy is negligibly ($\approx$3\%) lower
than that of G-SEQ for a fixed $\epsilon$ and varying~$n$ (Kronecker graphs).
The higher $\epsilon$ becomes, the more SC-OPT has advantage over G-SEQ. As
higher $\epsilon$ entails less circuit complexity (fewer substreams are
processed independently, assuming a fixed~$w_{max}$~\cite{crouch2014improved}),
\textbf{we conclude that the substream-centric MWM SC-OPT scheme provides
better approximation than G-SEQ when physical resources become more
constrained.}

\subsection{Influence of Blocking Parameter $K$}
\label{sec:perf-K}

\begin{figure}[h!]
%
	\centering
		\includegraphics[width=0.495\textwidth]{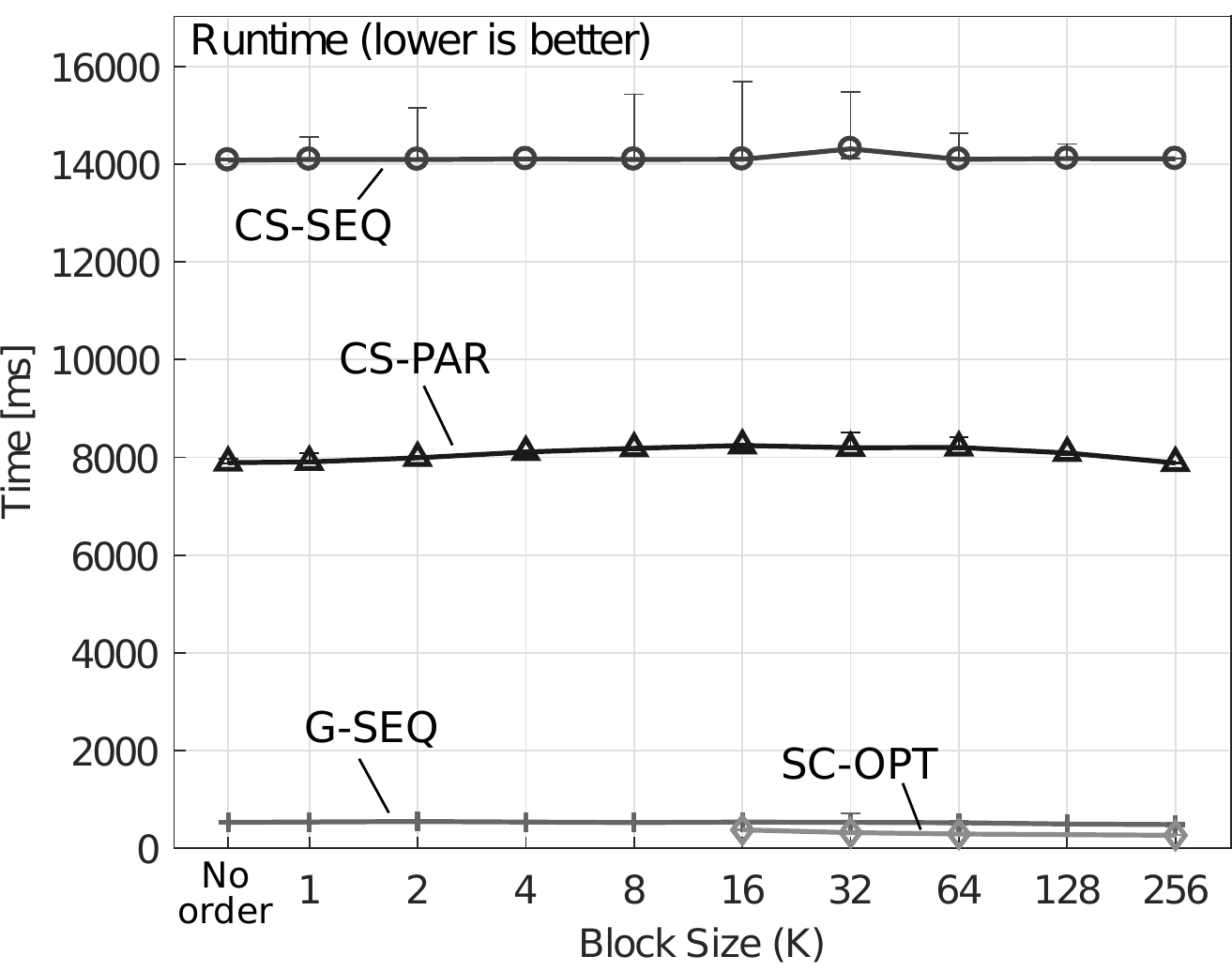}
		\includegraphics[width=0.48\textwidth]{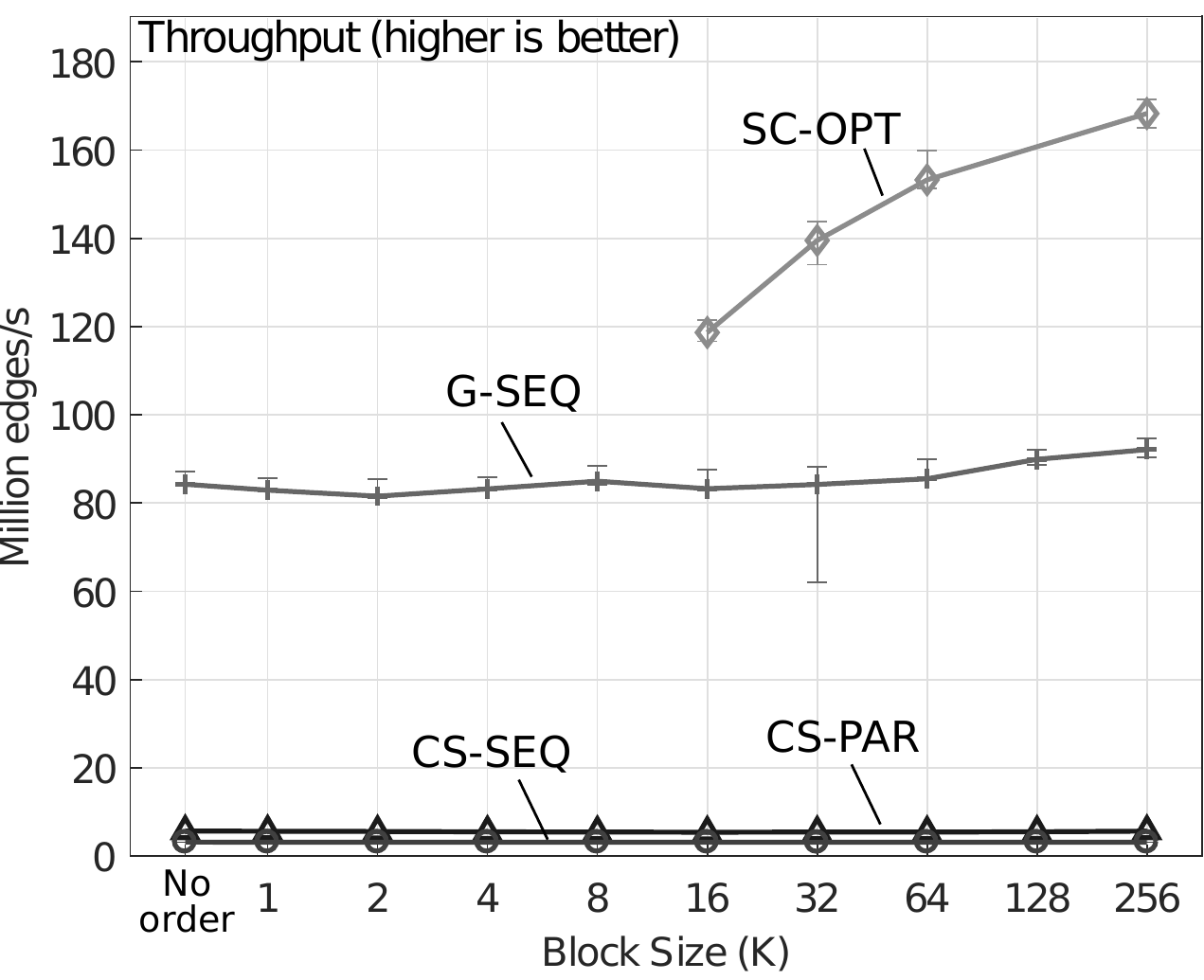}
%
	\caption{(\cref{sec:perf-K}) \textbf{Influence of epoch size $K$ on the performance}.
The input graph is Kronecker with $n = 2^{20}$.
$L=128$,
$T=4$, and $\epsilon = 0.1$.} 
\label{fig:perf-K}
		
\end{figure}

We also analyze the performance impact from $K$, a parameter that determines
how many rows in the streamed-in adjacency matrix are merged together using a
lexicographic ordering. Figure~\ref{fig:perf-K} illustrates the results.
%
%
%
On one hand, the CPU schemes cannot take significant advantage when $K$
increases, showing that no cache locality is exploited. On the other hand,
FPGA-based SC-OPT accelerates from $\approx$125M to $\approx$175M edges/s.
This is up to 2$\times$ faster than the work-optimal G-SEQ and up to 55$\times$
faster than CS-SEQ.  This is expected as the amount of stalling is reduced by a
factor of $n/K$.  Moreover, increasing $K$ allows to share more matching bits
between edges.  The performance impact is reduced when $K$ reaches 256. We
conjecture this is because of the random access to the matching bits, approaching
the peak random bandwidth.
%
%
%
Furthermore, G-SEQ outperforms all other CPU
implementations with up to $\approx$90M edges/s.  Compared to CS-SEQ
($\approx$3.15M edges/s) and CS-PAR ($\approx$5.6M edges/s), this is
$>$15$\times$.  Finally, parallelization comes with high overhead, such that
the four threads in CS-PAR achieve less than 2$\times$ speedup compared to
CS-SEQ.
%
%
\textbf{We conclude that our blocking scheme enables SC-OPT to achieve even
higher speedups.}

\subsection{Influence of Maximum Matching Count $L$}
\label{sec:perf-L}

Finally, we analyze the impact of $L$ on performance. $L$ is the
number of substreams and thus maximum matchings computed independently. 
CS-SEQ and CS-PAR achieve high performance with up to $\approx$400M edges/s for
$L=1$. The performance drops linearly with $L$ (X-axis has a logarithmic scale)
to $\approx$800k edges/s for CS-SEQ and $\approx$1.3M edges/s for CS-PAR. G-SEQ
also drops in performance as $L$ increases due to $\epsilon$ and $w_{max}$.
Since $L$ increases, we also increase the range of the weight
($L$ influences the approximation by $\epsilon = \sqrt[L]{w_{max}} -1$).
Thus, for $L=1$
the maximum edge weight is given by $w_{max} = 1$, allowing G-SEQ to drop many
edges in an early phase.  The drop of performance between $L=32$ and $L=64$ are
due to a change in $\epsilon$, requiring G-SEQ to store more data. Similarly,
we change $\epsilon$ between $L=128$ and $L=256$. 
\textbf{SC-OPT keeps its performance at $\approx$140M edges/s ($\approx$330ms)
and outperforms other schemes.}
%
%
%
%

\begin{figure}[h!]
%
	\centering
		\includegraphics[width=0.495\textwidth]{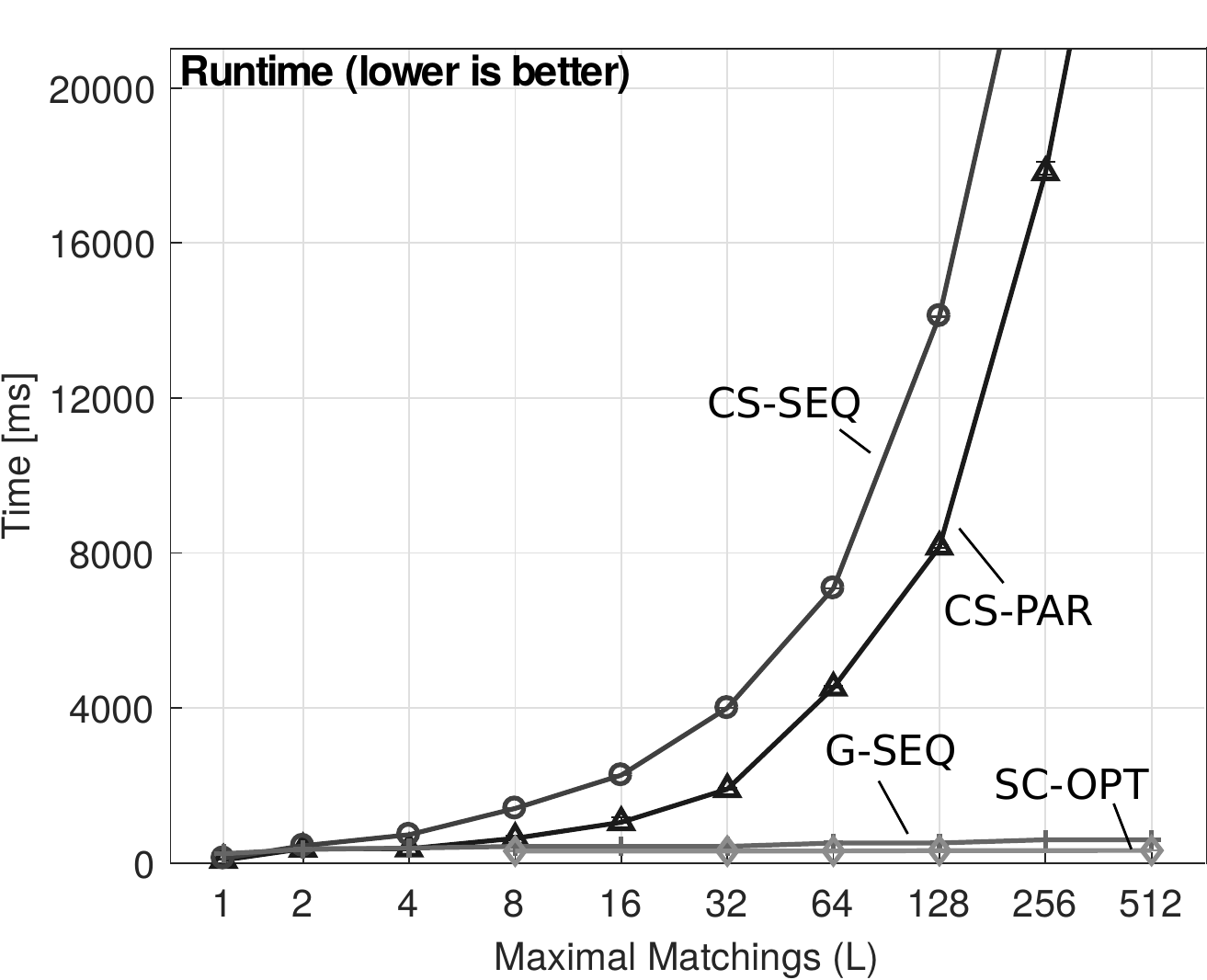}
		\includegraphics[width=0.48\textwidth]{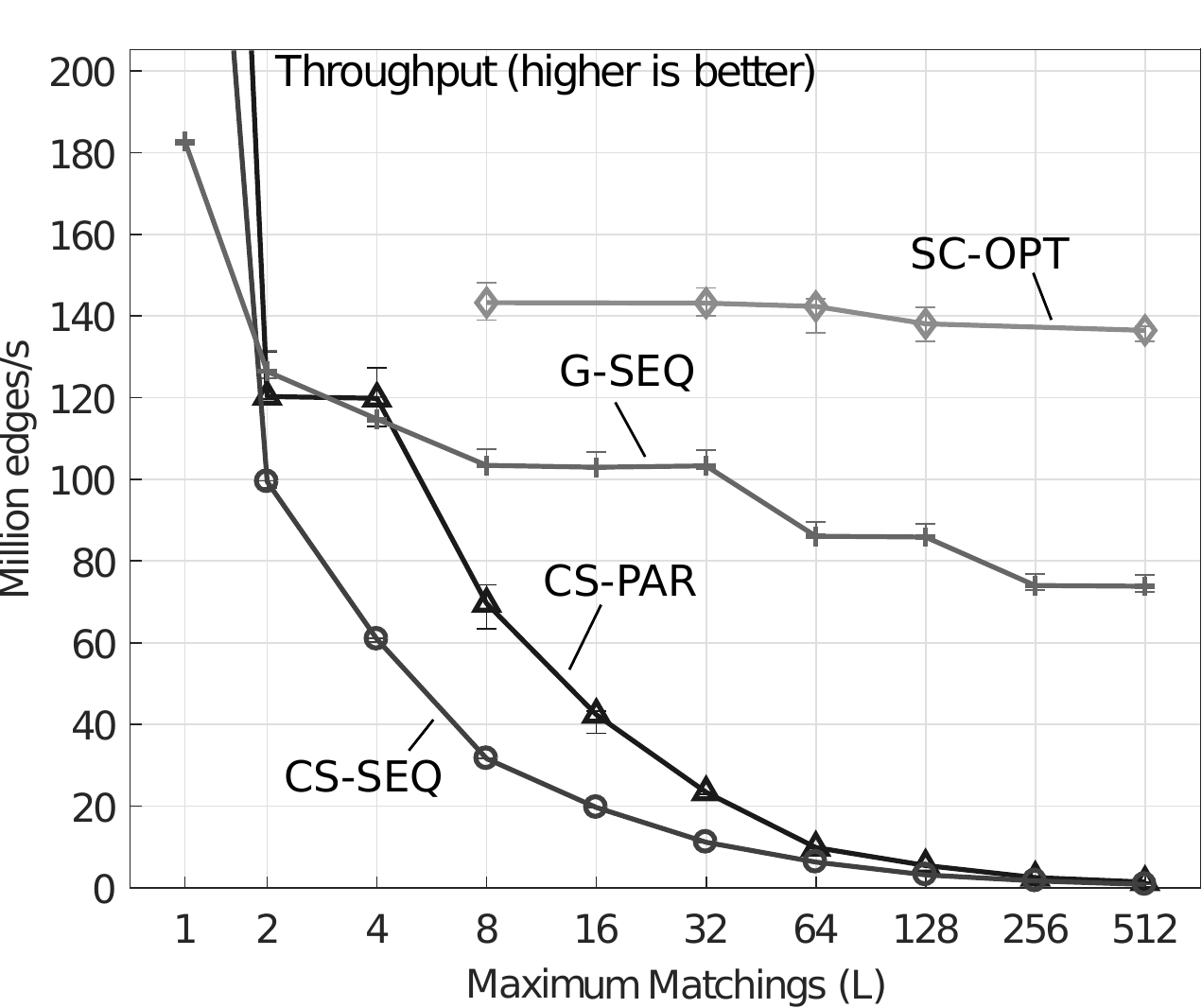}
%
\caption{(\cref{sec:perf-L}) \textbf{Influence of $L$ on performance}. 
The input graph is Kronecker ($n = 2^{20}, K=32, T=4$). As $L$ changes, $\epsilon$ changes
as follows: for $1 \leq L \leq 32$, we select $\epsilon = 0.6$, for $64\leq L
\leq 128$ we select $\epsilon = 0.1$, and for $256 \leq L \leq 512$ we select
$\epsilon = 0.03$; $w_{max}$ is given by $w_{max} = (1+\epsilon) ^L$.
We restricted the range of $L$ for SC-OPT due to the significant runtime
required to generate different bitstreams for evaluation.
}
%
\label{fig:perf-L}
\end{figure}

\subsection{FPGA Resource Utilization}
\label{sec:resource}

Table~\ref{table:evaluation:resources}
shows the usage of FPGA resources.
%
%
As maximum matchings are computed on the FPGA in one clock cycle, the number of
computed matchings~$L$ influences the amount of used logic. Moreover, for
SC-OPT, $K$ and $L$ determine the FPGA layout. Specifically, $K$ influences the
BRAM usage, since every element in the merging network requires two queues
which are each mapped to one BRAM unit.
We also consider the amount $B$ [bits] of BRAM allocated to storing the
matching bits. 
SC-OPT requires only 21\% of Arria-10's BRAM and 32\% out of
all ALMs for a design that outperforms other targets by at least $\approx$2$\times$
(Figures~\ref{fig:perf-K}--\ref{fig:perf-L}); these speedups can be
increased even further by maximizing circuitry utilization.
Finally, 
we include SC-SIMPLE in the analyses to illustrate the impact of additional
optimizations in SC-OPT.
CS-SIMPLE uses less ALMs (approx. 21\%) than CS-OPT (usage varies, up to 82\%).

\begin{table}[h!]%


\begin{minipage}{\columnwidth}
\begin{center}
\begin{tabular}{llll}
  \toprule
  \textbf{FPGA Algorithm} & \textbf{Parameters} & \textbf{Used BRAM} & \textbf{Used ALMs}  \\ \midrule 
  SC-SIMPLE & $\log B =12, L=8$ & 5.6 MBit (10\%) & 89,388 (21\%) \\ 
  SC-SIMPLE & $\log B=18, L=6$ & 21 MBit (38\%) & 88,920  (21\%) \\ 
  SC-OPT & $K=32, L=512$ & 11.5 MBit (21\%) & 151,998 (32\%) \\ 
  SC-OPT & $K=256, L=128$ & 24.8 MBit (45\%) & 350,556 (82\%) \\
 \bottomrule
\end{tabular}
\end{center}
\end{minipage}

\caption{(\cref{sec:resource}) \textbf{FPGA resource usage} for different
parameters.
%
%
}

\label{table:evaluation:resources}

\end{table}

%

\subsection{Energy Consumption}
\label{sec:energy}


We estimate the energy consumption of SC-SIMPLE and SC-OPT using the Altera
PowerPlay Power Analyzer Tool; see Table~\ref{table:evaluation:energy} {(we use
200MHz and include static power)}.
Furthermore, the host CPU (Broadwell Xeon E5-2680 v4) has TDP of 120
Watt~\cite{intelxeon} when all cores are in use {(We use TDP as the baseline
for the CPU because the utilized server is physically located elsewhere and we
  are unable to directly measure the used power)}. The TDP is an upper bound
  for CS-PAR at $T=64$.  \textbf{FPGA designs reduce consumed energy by
    \emph{at least} $\approx$88\% compared to the CPU.}

\begin{table}[h!]%


\begin{minipage}{\columnwidth}
\begin{center}
\begin{tabular}{lll}
  \toprule
  \textbf{Algorithm} & \textbf{Parameters} & \textbf{Energy Consumption [W]} \\ \midrule 
  SC-SIMPLE & $\log B = 18, L=6$ & 14.714 \\ 
  SC-SIMPLE & $\log B = 12, L = 8$ & 14.598 \\ 
  SC-OPT & $K=32, L=512$ & 14.789 \\ 
  SC-OPT & $K=256, L=128$ & 14.789  \\ 
  SC-OPT & $K=32, L=64$ & 14.657 \\ 
  CS-PAR & $T = 64$ & 120 \\
 \bottomrule
\end{tabular}
\end{center}
\end{minipage}

\caption{(\cref{sec:energy}) \textbf{Estimated energy consumption for different
parameters}.}
\label{table:evaluation:energy} \vspace{-1em}

\end{table}

%

\subsection{Design Space Exploration}
\label{sec:freq}

\begin{figure}[h!]
	\centering
	\includegraphics[width=0.65\columnwidth]{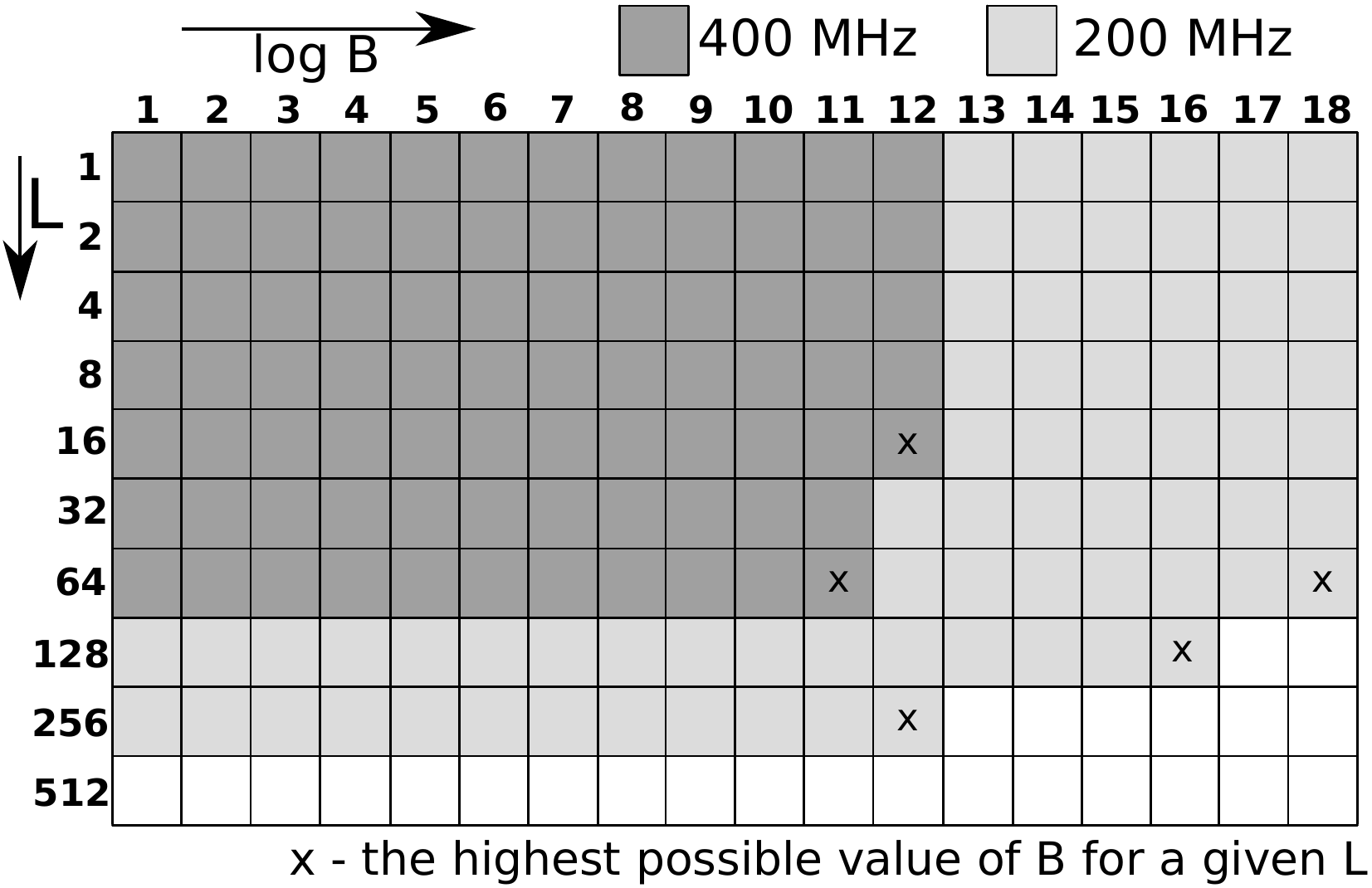}
  \caption{(\cref{sec:freq}) \textbf{Design space exploration: the used (available) frequencies}.} 
%
%
  \label{fig:evaluation:naive:frequency}
\end{figure}

We now briefly analyze the interaction between the performance of our FPGA
design and the limitations due to the clock frequency.
The resource usage, determined by $L$ and $B$, influences the
frequency upper bound due to wiring and logic complexity. We applied a grid
search to derive feasible frequencies for SC-SIMPLE; see
Figure~\ref{fig:evaluation:naive:frequency}
(we exclude SC-OPT as our analysis shown that the design is too
complex to run at 400MHZ  and we were only able to use it with 200MHz).
Dark grey indicates 400MHz, light
grey indicates 200MHz {(we use only the two frequencies as only those two
were supported by the Centaur framework at the time of evaluation)}. 
Two factors have shown to limit the performance. First, while
computing the matching, we use an addition with a variable that uses $L$ bits.
Thus, the addition complexity grows linearly with $L$. More importantly,
the BRAM signal propagation limits the frequency. For example, for
SC-SIMPLE and $\log B = 13$, the place and route report shows that the
reset signal to set all BRAM units to zero becomes the critical path.
{As alleviating these two issues would make our final design even more
complex, we leave it for future work. Specifically, we are now working on
a general FPGA substream centric engine that will feature pipelined
reset BRAM signals.}



%

\subsection{Optimality Analysis}

We also discuss how far the obtained results are from the maximum achievable
performance numbers; we focus on the most
optimized SC-OPT.
SC-OPT can process up to
$\approx$175M edges/s.  This is
close to the optimum due to different reasons: Firstly, the implementation can
process up to 1 edge per cycle (200M edges/s). Thus, the achieved
performance is optimal within only
$\approx$12\%.  Second, assuming that edges are read aligned from memory, it
allows {to read 8 edges per read request}. Further, if every edge requires its
own data chunk with matching bits, it needs 1 request per edge.  Overall, this
results in {$1+1/8 = 1.125$} read requests per edge. Under this assumption, the performance
is limited to {178M edges/s}. SC-OPT performs close to this bound, which is possible because \emph{the
matching bits can be shared between edges}.

\section{Beyond Substream-Centric MM}
\label{sec:beyond}

We now briefly discuss how to apply our substream-centric FPGA design to other
streaming graph algorithms.
%
%
%
First, we identify some MM schemes that also
divide the streamed dataset into substreams and can straightforwardly
be adopted to the hybrid CPU+FPGA system.
The \textbf{MWM algorithm by Grigorescu et al.~\cite{grigorescu2016streaming}}
reduces the MWM problem to $O(\epsilon^{-1} \log(n))$ instances of maximum
matchings, which could be processed on the FPGA analogously to our design; its
merging phase could also be executed on the CPU.  All our optimizations, such
as blocking, are applicable in this case.
Moreover, the \textbf{MWM algorithm by Feigenbaum et al.~\cite[Algorithm
4]{feigenbaum2005graph}} does not divide the stream of edges into substreams,
but its design would potentially allow for applying our blocking scheme.  A key
part of this algorithm is maintaining a certain value~$q_e$ associated with
each edge~$e$.  Given an edge $e = (u, v, w)$, $q_e$ depends on values $q_u$
and $q_v$ associated with vertices $u$ and $v$.  We can apply the blocking
pattern by storing $q_u$ for $u$ in BRAM, and streaming in $q_v$ for $v$.  
%
%
Next, the \textbf{MWM algorithm by Ghaffari~\cite{ghaffari2017space}} provides
a $(2+\epsilon)$-approximation. The algorithm compares the weights of incoming
edges to values $\phi$, indexed by $u$ or $v$. Therefore, it can be computed on
the FPGA using the blocking pattern by storing the values $\phi_u$ in BRAM, and
streaming $\phi_v$ from DRAM, similarly to matching bits in our design.
Further, as the algorithm requires postprocessing to derive the final result,
it could be also delegated to the CPU. 
%

We also identify algorithms unrelated to matching that could be enhanced with
our design.
The random \textbf{triangle counting algorithm by Buriol et
al.~\cite{buriol2006counting}} is also a suitable candidate for the presented
blocking pattern. The algorithm requires three passes. In pass~1, the number of
paths of length two in the input graph is computed. In pass~2, a random path of
length two is selected. In pass~3, the stream is searched for a certain edge,
dependent on the randomly selected path.  To reduce variance, passes~2--3 are
run in parallel using a pre-determined number of random variables (up to a
million). This also implies that in pass~3 every edge in the stream must be
checked against a million edges. To reduce the workload, {a hash map is used}. The
map is filled with edges which are expected to occur.
We propose the following approach to exploit the blocking pattern: the CPU
fills a hash map for each epoch with edges expected to arrive. The map is
passed to the FPGA. The edges for this epoch are streamed in and compared to
the pre-filled hash map. If the epoch changes, the next hash map is passed
over.
\section{Related Work}

Our work touches on various areas. We now discuss
related works, briefly summarizing the ones covered in previous sections
(streaming models in~\cref{sec:theory-analysis} and streaming maximum matching
algorithms in~\cref{sec:algs-theory}, Table~\ref{table:mcm},
and~\cref{sec:beyond}).

\subsection{Graph Processing on FPGAs}
The FPGA community has recently gained interest in processing
graphs~\cite{besta2019substream, besta2019graph, gianinazzi2018communication,
solomonik2017scaling, besta2017push, besta2017slimsell, besta2015accelerating,
besta2019slim, besta2018log, besta2019demystifying, besta2018survey, besta2019practice}
and other forms of general irregular computations~\cite{schweizer2015evaluating, besta2018slim, schmid2016high,
besta2014fault, gerstenberger2014enabling, tate2014programming, kepner2016mathematical,
di2019network, besta2015active}. First, some established CPU-related schemes
were ported to the FPGA setting, for example
vertex-centric~\cite{engelhardt2016vertex, engelhardt2016gravf},
GAS~\cite{zhou2017tunao}, edge-centric~\cite{zhou2017accelerating},
BSP~\cite{kapre2015custom}, and MapReduce~\cite{zhang2017boosting}.
There are also efforts independent of the above, such as
FPGP~\cite{dai2016fpgp}, ForeGraph~\cite{dai2017foregraph}, and
others~\cite{nurvitadhi2014graphgen, betkaoui2011framework, weisz2013graphgen,
zhou2016high, oguntebi2016graphops, kapre2015custom}. These works target
popular graph algorithms such as BFS or PageRank.
Multiplication of matrices and vectors~\cite{kwasniewski2019red,
besta2019communication} has also been addressed in the context of
FPGAs~\cite{licht2019flexible, sun2007sparse, zhuo2005sparse,
wang2007performance, elgindy2002sparse, dorrance2014scalable}; these efforts
could be used for energy-efficient and high-performance graph analytics on
FPGAs due to the possibility of expressing graph algorithms in the language of
linear algebra~\cite{kepner2016mathematical}.
\emph{{Our work differs from these designs as we focus on} the problem of
finding graph matchings. For more detailed analysis of related work, please
refer to Table~1}.


\subsection{Graph Matchings and FPGAs}
The only work related to matchings and FPGAs that we are aware of merely uses
matchings to enhance FPGA segmentation design~\cite{chang1998graph},
\emph{which is unrelated to deriving matchings and graph processing in
general}.

\subsection{Streaming Models and Algorithms}
We investigate the rich theory of streaming
models~\cite{henzinger1998computing, feigenbaum2005graph,
ahn2012graph, cormode2018independent, mcgregor2016better,
muthukrishnan2005data, datar2002maintaining,
chakrabarti2009annotations, aggarwal2004streaming, demetrescu2009trading,
karande2011online, dean2008mapreduce} and identify the semi-streaming
model~\cite{feigenbaum2005graph} as the best candidate for using together with
FPGAs to deliver algorithms with provable properties that match FPGA
characteristics such as limited memory.
We then investigate semi-streaming algorithms for maximum
matchings~\cite{feigenbaum2005graph, konrad2012maximum,
kapralov2014approximating, ahn2011linear, karp1990optimal,
goel2012communication, kapralov2013better, chitnis2016kernelization,
assadi2016maximum, mcgregor2005finding, zelke2012weighted, epstein2011improved,
grigorescu2016streaming, paz20172+, ghaffari2017space, crouch2014improved} and
identify the scheme by Crouch and Stubbs~\cite{crouch2014improved} that we use
as the basis for our substream-centric design \emph{that ensures
low-power, high-performance, and high-accuracy general maximum weighted
matchings on FPGAs}.

\subsection{Hybrid FPGA+CPU Platforms}
Finally, our work is related to the study of {hybrid CPU+FPGA
platforms~}\cite{owaida2017centaur, zhang2017boosting, andrews2004programming_missing, andrews2004programming, umuroglu2015hybrid, putnam2008chimps,
agron2006run, jidin2005extending, sidler2017accelerating, inta2012chimera, zhu2009buffer, zhu2009buffer}. We illustrate a case
study with maximum matchings and show that \emph{hybrid platforms can
outperform state-of-the-art parallel CPU designs in both performance and power
consumption}. {Other works on graph processing on hybrid FPGA-CPU systems
include a hybrid scheme for BFS~}\cite{umuroglu2015hybrid}.


\section{Conclusion}

An important problem in today's graph processing is developing high-performance
and energy-efficient algorithms for approximating maximum matchings.  Towards
this goal, we propose the first maximum matching algorithm for FPGAs. Our
algorithm is \emph{substream-centric}: the input stream is divided into
substreams that are processed independently on the FPGA and merged into the
final outcome on the CPU. This exposes parallelism while keeping communication
costs low: only $O(m)$ data must be streamed from DRAM to the FPGA.
Our algorithm is energy-efficient (88\% less consumed energy over a tuned CPU
variant) and provably accurate, fast (speedups of $>$4$\times$ over parallel
CPU baselines), and memory-efficient ($O(n \log^c n)$ required storage).

The underlying FPGA design uses several novel optimizations, such as merging
rows of the graph adjacency matrix and ordering resulting blocks
lexicographically. This enables low utilization of FPGA resources (only 21\%
of Arria-10's BRAM and 32\% out of all ALMs) while outperforming CPU baselines
by at least $\approx$2$\times$.  Both the FPGA implementation and the
substream-centric approach could be extended to other graph problems.

Finally, to the best of our knowledge, the proposed design is the first to
combine the theory of streaming with the FPGA setting.  Our insights coming
from the analysis of 14 streaming models and 28 streaming matching algorithms
can be used to develop more efficient FPGA designs.

\vspace{0.5em}
\sf
\macb{Acknowledgements}
We thank Mohsen Ghaffari for inspiring discussions
that helped us better understand graph streaming theory.
We also thank David Sidler for his help with the FPGA infrastructure.
Funded by the European 
Research Council (ERC) under the European Union's Horizon 2020 programme 
grant No. 678880.
TBN is supported by the ETH Zurich Postdoctoral Fellowship and Marie Curie Actions for People
COFUND program.




\renewcommand*{\bibfont}{\small}

{
\bibliographystyle{abbrv}
\bibliography{../thesis/references,sample-bibliography,survey-refs}
}

\end{document}